\documentclass[twocolumn,twocolappendix]{aastex63}
\graphicspath{{./}{figures/}}

\accepted{June 10, 2021}
\submitjournal{AJ}

\shorttitle{TASOC Photometry Pipeline}
\shortauthors{Handberg et al.}

\NewPageAfterKeywords

\usepackage{threeparttable}
\usepackage{booktabs}
\usepackage{fancyvrb}
\usepackage{placeins}

\newcommand{\TESS}{TESS\xspace} 
\newcommand{\fref}[1]{Figure~\ref{#1}}
\newcommand{\tref}[1]{Table~\ref{#1}}
\newcommand{\sref}[1]{Section~\ref{#1}}
\newcommand{\aref}[1]{Appendix~\ref{#1}}

\usepackage{rhmath}

\let\originaleqref\eqref
\renewcommand{\eqref}[1]{Equation~\originaleqref{#1}}

\newcommand{\Tmag}{\ensuremath{\mathrm{Tmag}}\xspace}
\newcommand{\Teff}{\ensuremath{T_\mathrm{eff}}\xspace}
\newcommand{\code}[1]{\texttt{#1}}

\newcommand{\SC}{$120$-sec\xspace}
\newcommand{\LC}{$1800$-sec\xspace}
\newcommand{\SCC}{$20$-sec\xspace}
\newcommand{\LCC}{$600$-sec\xspace}

\usepackage[nolist]{acronym}
\begin{acronym}[TASOC]
	\acro{TASOC}{TESS Asteroseismic Science Operations Center}
	\acro{TASC}{TESS Asteroseismic Science Consortium}
	\acro{TOI}{TESS Object of Interest}\acrodefplural{TOI}{TESS Objects of Interest}
	\acro{TIC}{TESS Input Catalog\acroextra{ \protect\citep{TIC}}}
	\acro{FFI}{Full Frame Image}
	\acro{TPF}{Target Pixel File}
	\acro{TBJD}{TESS Barycentric Julian Date}
	\acro{SPOC}{Science Processing Operations Center\acroextra{ \protect\citep{Jenkins2016}}}
	\acro{FOV}{field-of-view}
	\acro{WCS}{World Coordinate System}
	\acro{FITS}{Flexible Image Transport System}
\end{acronym}
\begin{document}
\title{\TESS Data for Asteroseismology: Photometry}

\correspondingauthor{Rasmus Handberg}
\email{rasmush@phys.au.dk}

\author[0000-0001-8725-4502]{Rasmus Handberg}
\affiliation{Stellar Astrophysics Centre, Department of Physics and Astronomy, Aarhus University, Ny Munkegade 120, DK-8000 Aarhus C, Denmark}

\author[0000-0001-9214-5642]{Mikkel N. Lund}
\affiliation{Stellar Astrophysics Centre, Department of Physics and Astronomy, Aarhus University, Ny Munkegade 120, DK-8000 Aarhus C, Denmark}

\author[0000-0002-6980-3392]{Timothy R. White}
\affiliation{Sydney Institute for Astronomy (SIfA), School of Physics, University of Sydney, 2006, Australia}
\affiliation{Research School of Astronomy and Astrophysics, Mount Stromlo Observatory, The Australian National University, Canberra, ACT 2611, Australia}
\affiliation{Stellar Astrophysics Centre, Department of Physics and Astronomy, Aarhus University, Ny Munkegade 120, DK-8000 Aarhus C, Denmark}

\author[0000-0002-0468-4775]{Oliver J. Hall}
\affil{European Space Agency (ESA), European Space Research and Technology Centre (ESTEC), Keplerlaan 1, 2201 AZ Noordwijk, The Netherlands}
\affiliation{School of Physics and Astronomy, University of Birmingham, Edgbaston, Birmingham B15 2TT, UK}
\affiliation{Stellar Astrophysics Centre, Department of Physics and Astronomy, Aarhus University, Ny Munkegade 120, DK-8000 Aarhus C, Denmark}

\author[0000-0002-1988-143X]{Derek~L.~Buzasi}
\affiliation{Department of Chemistry and Physics, Florida Gulf Coast University, 10501 FGCU Blvd. S., Fort Myers, FL 33965 USA}

\author[0000-0003-2595-9114]{Benjamin J. S. Pope}
\altaffiliation{NASA Sagan Fellow}
\affiliation{Center for Cosmology and Particle Physics, Department of Physics, New York University, 726 Broadway, New York, NY 10003, USA}
\affiliation{Center for Data Science, New York University, 60 Fifth Ave, New York, NY 10011, USA}

\author[0000-0002-4165-8592]{Jonas S. Hansen}
\affiliation{Stellar Astrophysics Centre, Department of Physics and Astronomy, Aarhus University, Ny Munkegade 120, DK-8000 Aarhus C, Denmark}

\author[0000-0002-6956-1725]{Carolina von Essen}
\affiliation{Stellar Astrophysics Centre, Department of Physics and Astronomy, Aarhus University, Ny Munkegade 120, DK-8000 Aarhus C, Denmark}
\affiliation{Astronomical Observatory, Institute of Theoretical Physics and Astronomy, Vilnius University, Sauletekio av. 3, 10257, Vilnius, Lithuania}

\author[0000-0003-1001-5137]{Lindsey Carboneau}
\affiliation{Department of Chemistry and Physics, Florida Gulf Coast University, 10501 FGCU Blvd. S., Fort Myers, FL 33965 USA}
\affiliation{School of Physics and Astronomy, University of Birmingham, Edgbaston, Birmingham B15 2TT, UK}
\affil{Stellar Astrophysics Centre, Department of Physics and Astronomy, Aarhus University, Ny Munkegade 120, DK-8000 Aarhus C, Denmark}

\author[0000-0001-8832-4488]{Daniel Huber}
\affiliation{Institute for Astronomy, University of Hawai`i, 2680 Woodlawn Drive, Honolulu, HI 96822, USA}

\author[0000-0001-6763-6562]{Roland~K.~Vanderspek}
\affiliation{Department of Physics and Kavli Institute for Astrophysics and Space Research, MIT, Cambridge, MA 02139, USA}

\author[0000-0002-9113-7162]{Michael~M.~Fausnaug}
\affiliation{Department of Physics and Kavli Institute for Astrophysics and Space Research, MIT, Cambridge, MA 02139, USA}

\author[0000-0002-1949-4720]{Peter~Tenenbaum}
\affiliation{NASA Ames Research Center, Moffett Field, CA 94035, USA}
\affiliation{SETI Institute, Mountain View, CA 94043, USA}

\author[0000-0002-4715-9460]{Jon~M.~Jenkins}
\affiliation{NASA Ames Research Center, Moffett Field, CA 94035, USA}

\author{and the T'DA Collaboration}


\begin{abstract}
Over the last two decades, asteroseismology has increasingly proven to be the observational tool of choice for the study of stellar physics, aided by the high quality of data available from space-based missions such as CoRoT, \Kepler, K2 and \TESS. \TESS in particular will produce more than an order of magnitude more such data than has ever been available before.

While the standard \TESS mission products include light curves from \SC observations suitable for both exoplanet and asteroseismic studies, they do not include light curves for the vastly larger number of targets observed by the mission at a longer \LC cadence in \acp{FFI}. To address this lack, the TESS Data for Asteroseismology (T'DA) group under the TESS Asteroseismic Science Consortium (TASC), has constructed an open-source pipeline focused on producing light curves for all stars 
observed by TESS at all cadences, currently including stars down to a TESS magnitude of 15. The pipeline includes target identification, background estimation and removal, correction of FFI timestamps, and a range of potential photometric extraction methodologies, though aperture photometry is currently the default approach. For the brightest targets, we transparently apply a halo photometry algorithm to construct a calibrated light curve from unsaturated pixels in the image.

In this paper, we describe in detail the algorithms, functionality, and products of this pipeline, and summarize the noise metrics for the light curves. Companion papers will address the removal of systematic noise sources from our light curves, and a stellar variability classification from these. 
\end{abstract}
\keywords{Asteroseismology, Variable stars, Astronomy data analysis, CCD photometry}

\enlargethispage{1.0\baselineskip}


\section{Introduction}\label{sec:intro}
Asteroseismology, the study of stellar oscillations, has proven itself to be a powerful tool for accurate determination of global stellar properties like mass, radius and age, utilized \eg in exoplanet characterization \citep[see \eg][]{Batalha2011}, and for studies of the galaxy through galactic archaeology \citep[see \eg][]{Miglio2013}. Furthermore, in its own right, asteroseismology enables unique studies of the detailed internal structure and evolution of stars across the Hertzsprung-Russell diagram (HRD) \citep[see \eg][]{TheBook}.

During the last couple of decades asteroseismic investigations have undergone rapid development, fueled to a large extent by the large amount of high-quality data available. Starting with bespoke studies of single stars from the ground and space to large photometric surveys of thousands of stars from space-based missions like CoRoT, \Kepler and K2 \citep{Baglin2002,Borucki2010,Howell2014}, the move towards larger data sets has naturally meant a shift to a more pipeline-oriented processing of the data \citep[see \eg][]{Jenkins2017,kasoc_filt,K2P2,Vanderburg2014,Luger2016}. Since stars exhibiting asteroseismic oscillations span the HRD, the oscillation signals themselves span orders of magnitudes in both frequencies and amplitudes, which on its own is a challenge for any processing.

With data from the Transiting Exoplanet Survey Satellite \citep[\TESS; ][]{Ricker2014,TESSInstrumentHandbook} the asteroseismic community is faced with a new set of challenges concerning the preparation of data.
\TESS gathers \acp{FFI} of $96\degree\!\times\!24\degree$ regions of the sky at \LC cadence, but no photometric extractions are performed, and hence no light curves are released as official data products. Existing community-built FFI photometry software includes \code{Eleanor} \citep{eleanor}, the MIT Quick-look pipeline \citep{quicklookpipeline,Huang2020a}, the SPOC FFI pipeline \citep{Caldwell2020}, as well as difference imaging \citep{Oelkers2018,cdips,Montalto2020} and PSF methods \citep{Nardiello2019}. However, neither of these techniques are optimized for asteroseismology, which is a large science driver for TESS photometry.

Therefore, the ``TESS Data for Asteroseismology'' (T'DA) coordinated activity within the \TESS Asteroseismic Science Consortium \citep[TASC; ][]{Lund_Azores2017} has been tasked with delivering light curves ready for asteroseismic analysis
for each target observed by TESS, including those in the \acp{FFI} and from \acp{TPF}. This is achieved using a custom-built, Open Source, pipeline the first steps of which we describe in this paper. Data from the pipeline are made available to the community via the data base of the \TESS Asteroseismic Science Operations Center (TASOC\footnote{\url{https://tasoc.dk}}) and the Mikulski Archive for Space Telescopes (MAST\footnote{\url{https://archive.stsci.edu/hlsp/tasoc}}).
This collaborative effort is building on the combined experience in the TASC community gained from previous missions, various types of oscillating stars, and different analysis methods, to process the \TESS data in a way which yields data optimized for detailed scientific investigations in all sub-fields of asteroseismology.

We note that while the ``TASOC pipeline'' is developed with a focus on asteroseismology, the data will also be useful for, \eg, exoplanet science, eclipsing binaries, and the study of stellar activity (rotation, flares, \etc).

\section{The TASOC Pipeline}
The TASOC pipeline is developed by the T'DA group primarily to serve the asteroseismic community of TASC. Specifically, T'DA is responsible for delivering (1) raw photometric time series from \acp{FFI} (\LC cadence) and \acp{TPF} (\SC cadence), (2) light curves corrected for systematic signals, ready for asteroseismic analysis, and (3) a classification of the variability/oscillation/pulsation type of each light curve from (1)-(2) --- enabling each of the working groups of TASC, which are organized according to stellar type, to better target their analysis. We note that from TESS Cycle~3 onwards a new FFI cadence of \LCC will be adopted, and a new \SCC cadence data product will be introduced. The adoption of these new cadences poses no challenge for the pipeline.

The full TASOC pipeline is open-source and available on GitHub\footnote{\url{https://github.com/tasoc}}. All data products from the pipeline are available via the TASOC database and as a ``high-level science product'' (HLSP) on MAST, see \sref{sec:data_formats} for more details.

\begin{figure*}
    \centering
    \includegraphics[width=0.7\textwidth]{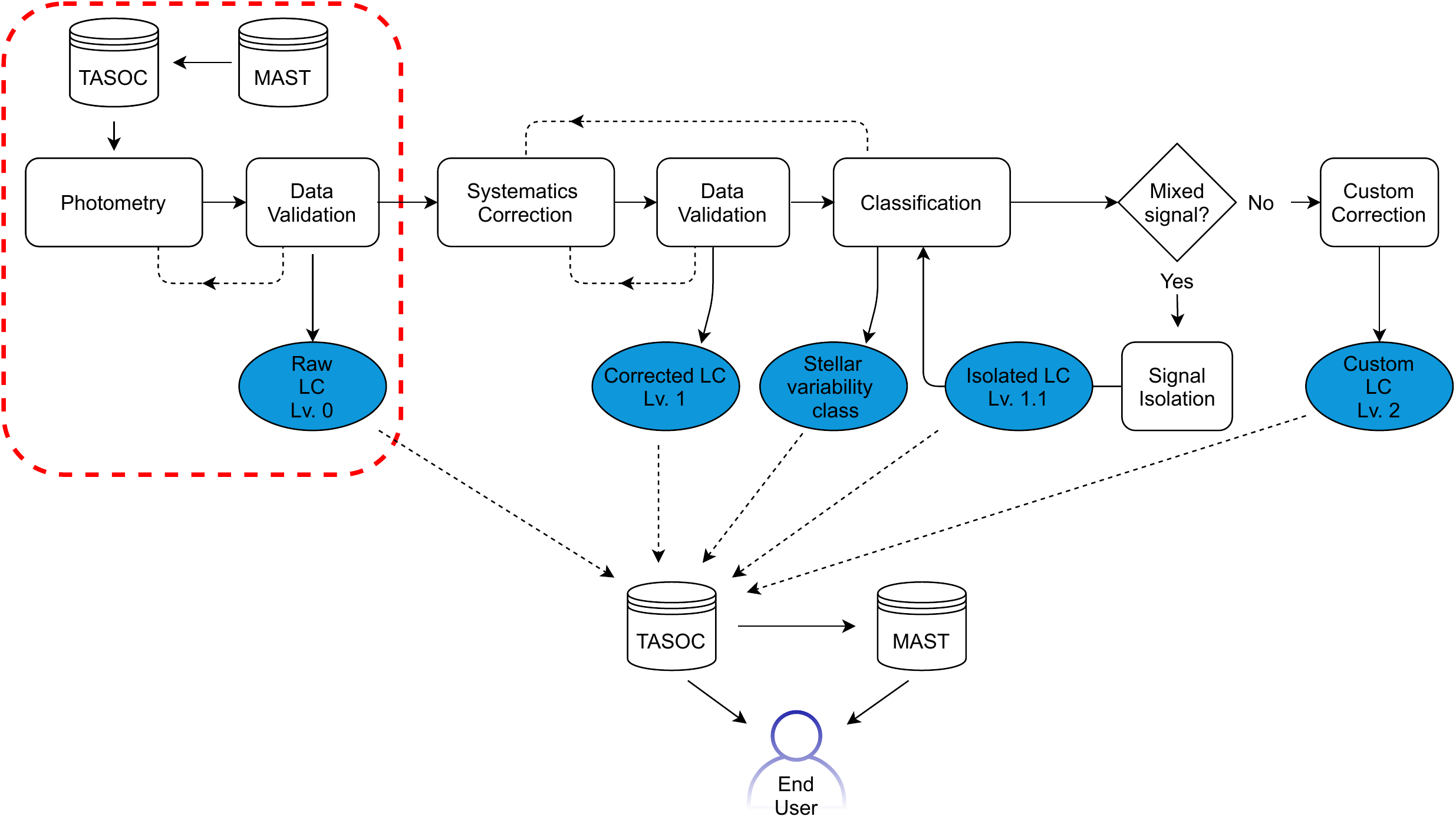}
    \caption{The overall structure of the full TASOC pipeline, with modules given as rectangular boxes, data products as ellipses, and ``TASOC'' and ``MAST'' indicate the databases hosting the data products. Dashed lines between modules indicate that an iteration might take place. The part enclosed by the red dashed line indicates the pipeline component described in this paper. The ``systematics correction'' part of the pipeline is described in Lund \etal (in prep.), while the ``classification'' is detailed in Audenaert \etal (in prep.).}
    \label{fig:pipeline}
\end{figure*}

The first step in the TASOC pipeline is photometric extraction of light curves for every target that falls on a \TESS \ac{FFI} or \ac{TPF}, down to a \TESS magnitude (\Tmag) of 15 -- this limit is set to reduce the processing time and because targets of interest to asteroseismology are typically brighter, but could easily be changed if fainter targets are requested. This first step will be the focus of this paper.
The following steps in the pipeline include the correction of systematics, described in Lund \etal (in prep.), and stellar classification, described in Audenaert \etal (in prep.). \fref{fig:pipeline} gives a general overview of the TASOC pipeline, and the different data products produced. The red dashed line encloses the component parts described in this paper.

One very important design goal of the whole TASOC pipeline is that \emph{all processing must be achievable in no more than 27 days of runtime}. This is due to the simple fact that the length of a single \TESS observing Sector is 27 days, and we need to be able to process one Sectors worth of data fully before the next Sector arrives. This necessitates the constraint that all the methods employed should not be unnecessarily computationally intensive. In any case, the whole pipeline has intentionally been constructed to be easily scaled up to larger computer systems should the need arise and more computational intensive methods have to be employed. In that case, to limit total processing time the underlying computer hardware could simply be expanded without significant changes to the pipeline described here.

\section{The Photometry Pipeline}
\begin{figure}
    \centering
    \includegraphics[width=\columnwidth]{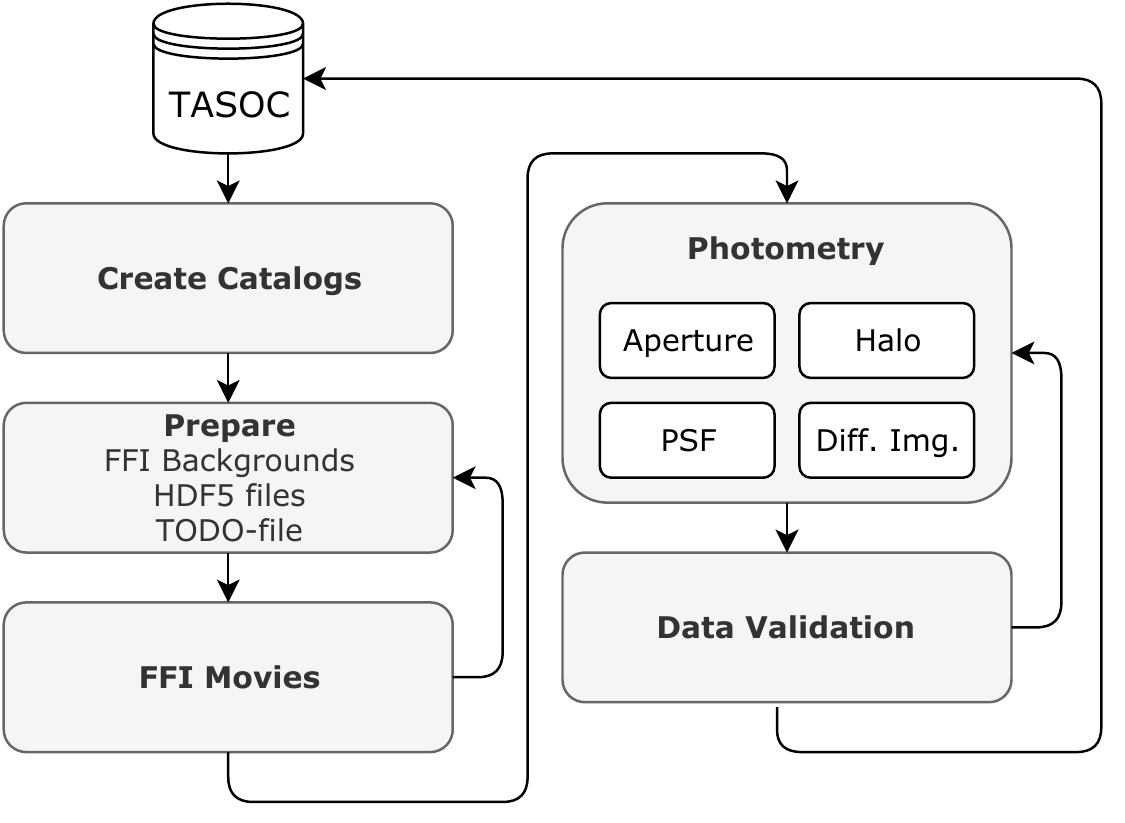}
    \caption{Structure of the TASOC Photometry pipeline.}
    \label{fig:photometry_pipeline}
\end{figure}
The goal of the photometry pipeline is to extract light curves for all targets observed by \TESS down to \Tmag$\leq15$. \TESS is producing pixel-level data in two distinct modes: \acp{FFI} and \acp{TPF}. The first, as the name suggests, are the full $2048\times2048$ pixel images for each of the 16 CCDs, stacked to \LC cadence (the base integration of data in TESS is 2 seconds). The \acp{TPF} consist of small postage-stamps around pre-selected stars, usually only a few tens of pixels on each side, stacked to \SC cadence.
The basic input for the TASOC photometry pipeline are the calibrated \TESS \acp{FFI} and \acp{TPF}, which are FITS (\acl{FITS}) files containing the individual pixels collected by \TESS as a function of time. The basic pixel-level calibrations (\eg dark, smear, flatfield) are done by the \TESS Science Processing Operations Center \citep[SPOC;][]{Jenkins2016}, which means that each pixel we use as input has units of counts per second and therefore instrumental effects at the detector level will ideally have been sufficiently calibrated. They also contain a \ac{WCS} solution, which maps the pixel-coordinates onto sky-coordinates (right ascension and declination). In all our processing we assume that these calibrations are sufficient, with the notable exception of the very brightest stars (see \sref{sec:bright_stars}).

The TASOC Photometry pipeline is split into a few separate components as shown in \fref{fig:photometry_pipeline}, which are run in succession. The three main components are ``prepare'', ``photometry'', and ``data validation''. 
We will go into the details of each of these in the following sections, but in general terms the process can be outlined as follows:

\paragraph{Prepare} 
The first step in the processing is to prepare the basic input files in such a way that the following photometric extraction of individual targets can be easily parallelized. The prime objective of this step is to perform all operations that need to have the information from the full FFIs available, \eg, estimate the background flux and flag problematic pixels (see \sref{subsec:background_estimation}), and to restructure the FITS files into a set of HDF5 files (Hierarchical Data Format). Among other things, the HDF5 file format allows us to extract arbitrary sub-sections of the \acp{FFI} without having to load the entire \ac{FFI} into memory.

\paragraph{Photometry}
The primary task is then to extract the brightness as a function of time for each star in the \acp{FFI} and \acp{TPF}, together with other relevant quantities, from the pixel-level data. This can be done in many different ways, several of which are implemented in the pipeline. The method that is used for the majority of stars in the TASOC pipeline is simple aperture photometry (SAP), but other methods are also used in special cases. The setup also allows us to easily add additional methods in the future. See \sref{sec:photometry} for details.

\paragraph{Data Validation}
The objective of the data validation module is to check the final results of the photometry and the generated light curves. Various diagnostic metrics are checked to filter out erroneous light curves, so these will not be propagated to the following steps of the overall TASOC pipeline or released to the community. It will also flag spurious light curves, which are deemed of sufficient quality to avoid rejection, but require warning users that potential issues may exist. See \sref{sec:validate} for details.

\subsection{Catalogs and TODO-files}\label{sec:catalogs_and_todo}
In \TESS (unlike \Kepler/K2) we can safely assume that we have a very good catalog of all stars of interest in the sky, due to \TESS targets usually being much brighter. This catalog is provided in the form of the TESS Input Catalog \citep[TIC; ][]{Oelkers2016,Stassun2018,Stassun2019}, which in its most recent incarnation (TIC-8) is based on the Gaia DR2 \citep{GaiaDR2} and 2MASS \citep{2mass} catalogs.
TASOC is storing a full copy of the TIC in a dedicated PostgreSQL database.

The first step in the photometry pipeline is to create full catalogs of all stars known to fall on or near the \TESS detectors during a given observing Sector, to be used in the following steps in the pipeline. These catalogs are copied directly from the TIC, but will only contain targets relevant for the observing Sector and CCD in question, and will be stored into local SQLite files. SQLite is a lightweight relational database management system that has the advantage of not requiring a database-server engine running, but is simply embedded into a file-system. Storing the catalogs into SQLite files gives the advantage that the program can run off-line from the main TASOC database systems, and will not be dependent on the response time of such a server or the network connecting to it.

In order to generate a catalog-file for a given \TESS observing Sector and CCD, we query the TIC using Q3C \citep{Koposov2006} for all targets falling within the footprint of the CCD with an additional 0.1\degree\ buffer around. We filter out any targets marked, \eg, as duplicates, and extract only the most relevant information to be stored into the SQLite files. This includes target position in RA and DEC (J2000), proper motions, \TESS magnitudes (\Tmag), and effective temperature (\Teff). We furthermore calculate the positions of the stars at the \emph{reference time} of the Sector (usually around apogee of the first orbit of the Sector) by projecting the coordinates forward in time with the known proper motions. These coordinates are the ones used in the rest of the pipeline as the reference coordinates of the targets.
For one full \TESS observing Sector a total of 16 catalog files are generated, one for each CCD.

Once we have created full catalogs of all targets, we can then create the list of targets to be processed in the photometry, as well as the following steps in the full TASOC pipeline. This file is called the TODO-file and, like the catalogs, this is also stored in an SQLite file. In this initial step, only the primary table of targets to be processed will be created, but at each step in the TASOC pipeline more tables will be added to the TODO-file, containing results and diagnostic information. They are therefore the central repository for information in the TASOC pipeline.

For FFIs this means simply taking all targets that are brighter than the set threshold $\Tmag < 15$, and adding them to the TODO-list if they fall within the FOV of one of the cameras at the reference-time of the Sector. For this we use the \ac{WCS} transformation provided in the header of the \acp{FFI}, including the Simple Imaging Polynomial (SIP) image distortion correction.

For each \ac{TPF} we add the central primary target to the list, but furthermore also search for secondary targets, identified as other targets brighter than the cutoff which fall within the observed pixels in the postage-stamp. If any such targets are found, these are added to the list. After all \acp{TPF} have been processed in this way, we filter the list by removing any secondary targets which were also identified as primary targets. We also check secondary targets which were found to be present in more than one \ac{TPF}, only keeping the one that is the furthest distance from the postage-stamp edge, since this is more likely to have captured more of the flux from the secondary target and not be truncated by the edge.

\subsection{Background estimation and quality flags}\label{subsec:background_estimation}
\begin{figure*}
    \centering
    \begin{interactive}{animation}{sector001_camera1_ccd2.mp4}
        \includegraphics[width=\textwidth]{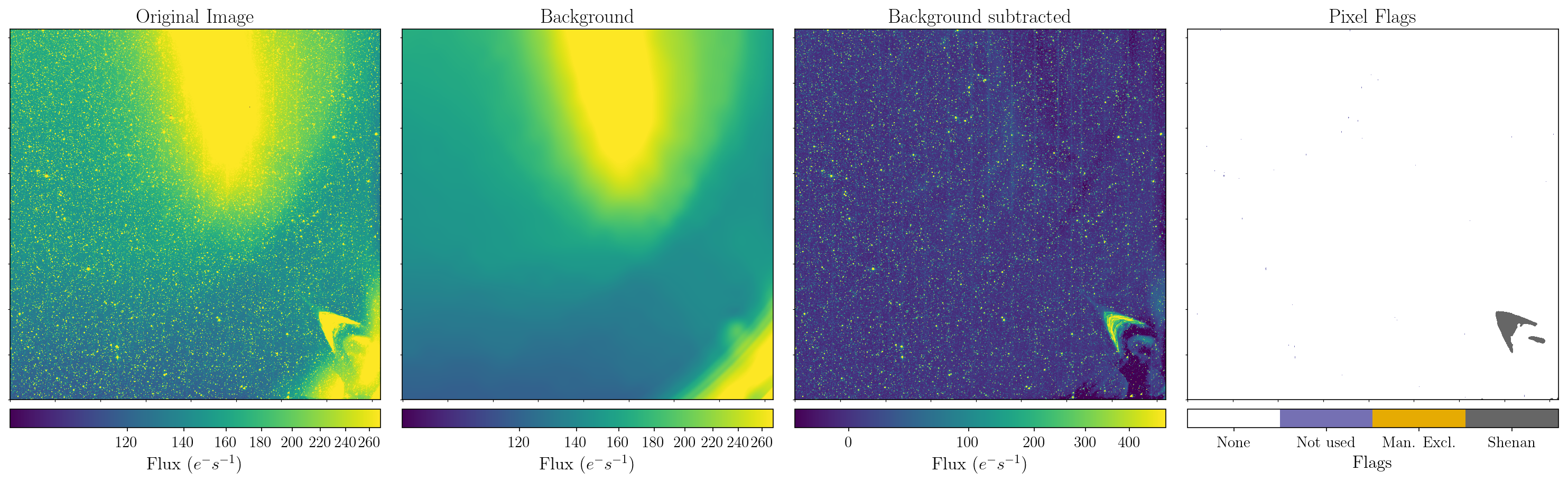}
    \end{interactive}
    \caption{Background extraction in FFIs, here for Sector 1, Camera 1, CCD 2. Panel 1 (leftmost): the raw calibrated FFI image; Panel 2: the estimated background including both the broad background and the corner glow; Panel 3: the background-corrected image; Panel 4 (rightmost): identification and flagging of the residual lens-flare ``wineglass'' feature in the background-corrected image. An animated version of this figure is available online, showing the change of the background during the full Sector [duration: 01:25].}
    \label{fig:backgrounds}
\end{figure*}

While \acp{FFI} provided by the \TESS mission have been calibrated, nothing has been done to remove sky background contributions to the images, originating from, \eg, scattered light from the Earth, Moon, other Solar-system objects and Zodiacal light. For \acp{FFI}, the first step in the processing is therefore to estimate the sky background level in each image. The challenge here is to estimate the changing large-scale background structures without including effects from variable stars (See \fref{fig:backgrounds}).

The first step is to mask out pixels in each image that are invalid or where very bright stars are present. Currently this is done simply by manually excluding regions of pixels with known problems (\eg Mars in Sector 1) and making a threshold in the flux above which pixels are masked out. We do have plans in the future to use the known locations of bright stars from the catalog to improve the masking process, but this is not yet fully implemented.

The full $2048\!\times\!2048$ pixel image is then cut into $64\!\times\!64$ pixel tiles, wherein a $3\sigma$ sigma clipping from the median value is performed to further mask out stars. The mode of the image is then estimated using the same prescription as is used by SExtractor \citep[SE;][]{SExtractor}:
\begin{equation}\label{eq:background}
\footnotesize 
    \mathrm{mode} = \begin{cases}
        2.5\cdot\mathrm{median} - 1.5\cdot\mathrm{mean}, & \frac{\mathrm{mean} - \mathrm{median}}{\mathrm{std}} \leq 0.3 \\
        \mathrm{median}               , & \text{otherwise}
    \end{cases}%
\end{equation}
These values are then run through a $3\!\times\!3$ moving median filter and interpolated back onto the full image size. Other methods for estimating the sky backgrounds were also tested, and we concluded that the method described above performed the best in terms of both recovering the background and computational speed. See \aref{app:background_tests} for more details.

Another dominant feature of the \TESS FFIs is the so-called ``corner glow'' \citep{TESSInstrumentHandbook} which is believed to be caused by stray light from stars outside the camera \ac{FOV}. The effect manifests itself in the four corners of the camera \ac{FOV} farthest away from the camera center (in one corner per CCD) as a rise in the background level of $\sim$1--2 times the overall sky background (see \fref{fig:backgrounds}). Since this feature is only present in the corner of the CCD and seems primarily to be a function of the radial distance from the camera center, the normal method using square tiles to estimate the background is sub-optimal in capturing the corner glow.
We mitigate this by re-parameterizing the image by calculating the Euclidean distance from each pixel to the camera center. We then bin the pixels into 15 pixels wide bins starting from a distance of 2400 pixels and upwards, and in each bin we estimate the background in the same way as described above. This effectively creates ``radial tiles'' near the corner, where the background is estimated from pixels having the same radial distance. The bins are then interpolated back onto the full image and we are left with a background profile which rises smoothly towards the corner.

Our final estimate of the background for a single \ac{FFI} will be a combination of the square tiles and radial components. We iteratively isolate the two components and create the final background from their sum. For a given \ac{FFI}, $\mathcal{I}$, the two background components are iterated three times, first estimating the radial component,  $\mathcal{B}_\mathrm{radial}$, from $\mathcal{I} - \mathcal{B}_\mathrm{square}$, and then estimating the square component, $\mathcal{B}_\mathrm{square}$, from $\mathcal{I} - \mathcal{B}_\mathrm{radial}$. The final background is $\mathcal{B} = \mathcal{B}_\mathrm{square} + \mathcal{B}_\mathrm{radial}$.
The contribution from the radial component, compared to the pure square component, varies across the Sectors with the level of scattered light, raising the average estimated background level in the corners on the order of a couple of percent. In cases with additional features near the corners (\eg in \fref{fig:backgrounds}) the largest differences can go up to $\sim25\%$, due to the improved separation of the radial rise and non-radial contributions.

Once the background has been estimated for all images for a given CCD, the final background images are constructed where the estimated background for each pixel is smoothed in time using using a 3-point (1.5 hour) moving average filter. This avoids very rapid changes in the background which could introduce high-frequency noise in the final photometry. At the same time, we construct an average image for each CCD, which is the co-added images after subtracting the final background, divided with the number of images. This sum-image is used in some of the following steps for locating stars and is also included in the final output files.

Once we have constructed the background-subtracted \acp{FFI}, we also want to flag any areas of the images affected by additional features that are too sharp to be captured by the background estimation described above. This includes \eg the lens-flares seen in Sector~1 because of Mars entering the \ac{FOV} of camera 4 (see \fref{fig:backgrounds}). These features move across the images and will impact different locations of the focal plane at different times. The goal is therefore to flag these features in each image by locating any pixels affected, and propagate this information to the final light curves.

After estimating the the backgrounds and average image as described above, we subtract the average image from each individual background-subtracted \ac{FFI}, creating a difference image showing the departures from the ``norm''. To avoid flagging variable pixels with small spatial scale structures, \eg variable stars, the difference image is run through a $15\!\times\!15$ pixel median filter. Any pixels departing more than a set threshold (nominally ${\sim}40$ counts per second) are flagged as irregular.
As the final outcome of this process we produce a set of pixel-level flag-images, which for each pixel and for each timestamp contains flags indicating the state of that pixel. This includes if the pixel was manually excluded, otherwise not used in the estimation of the background or if it has been flagged using the procedure described above. These pixel flag images are also stored internally in the HDF5 files for the given camera.

The next step in the ``prepare'' module is to do a matching between the data quality flags provided by the \TESS team in the \acp{TPF} and \acp{FFI} \citep[see][]{Twicken2020}, because the quality flags of the FFIs are generally not populated. For each CCD a small number of \acp{TPF} belonging to that CCD are randomly chosen and their quality flags are binned onto the \ac{FFI} cadence by enabling the flag if it is set in {\em any} of the corresponding TPF cadences.
We then compare these quality flags from the TPFs with the ones provided with the FFIs, only including flags relevant for all targets (\eg spacecraft attitude tweaks, course pointing, momentum dumps, \etc), and propagate any missing flags to the FFIs.
We do not include the ``Manual Exclude'' flag, since a single \SC data point marked as bad does not necessarily mean that the combined \LC FFI should be rejected. We also found that including this would discard ${\sim}20\%$ of all data in Sector 1.
This step is included to ensure that quality flags present in the TPFs are propagated into the lower-level FFIs. In most cases, particularly in later Sectors, this has no effect on the final quality stamps, as the FFI quality stamps have been populated correctly.

The final step in the ``prepare'' module is to create movies as a function of time of the original FFI images, estimated backgrounds, image-subtracted FFIs and extracted pixel-level flags, as shown in \fref{fig:backgrounds}.
These are used for visual inspection of the performance of the background estimation and the pixel-level flags. In cases where individual pixels need to be removed due to instrumental effects not identified in the automatic processing these would be added and the ``prepare'' module would be re-run, using the new manually excluded pixels.

\subsection{Timestamps of FFIs}\label{sec:timestamps}
The FFIs provided by \TESS on the MAST website all contain a timestamp in the header of each image. These timestamps are provided in \ac{TBJD}, and might naively be anticipated to be of sufficient quality to use in any light curves that one would extract from the FFIs. However, these timestamps have been calculated for a fixed position in the center of the images, but since the barycentric correction depends on the exact coordinates of the target in question and each camera spans several degrees on sky, these timestamps actually need to be be re-corrected to the proper coordinates of the star being extracted.

We start by reverting the barycentric correction of the provided FFI timestamps back to Julian Dates in TDB (Barycentric Dynamical Time) in the \TESS frame of reference.
We then use SPICE kernels provided by the \TESS Team to calculate the accurate position of the \TESS spacecraft at the time of observation and calculate the light-travel time correction resulting from the different light-paths from the position of the target to the TESS spacecraft versus the Solar System barycenter.
\begin{figure}
    \centering
    \includegraphics[width=\columnwidth]{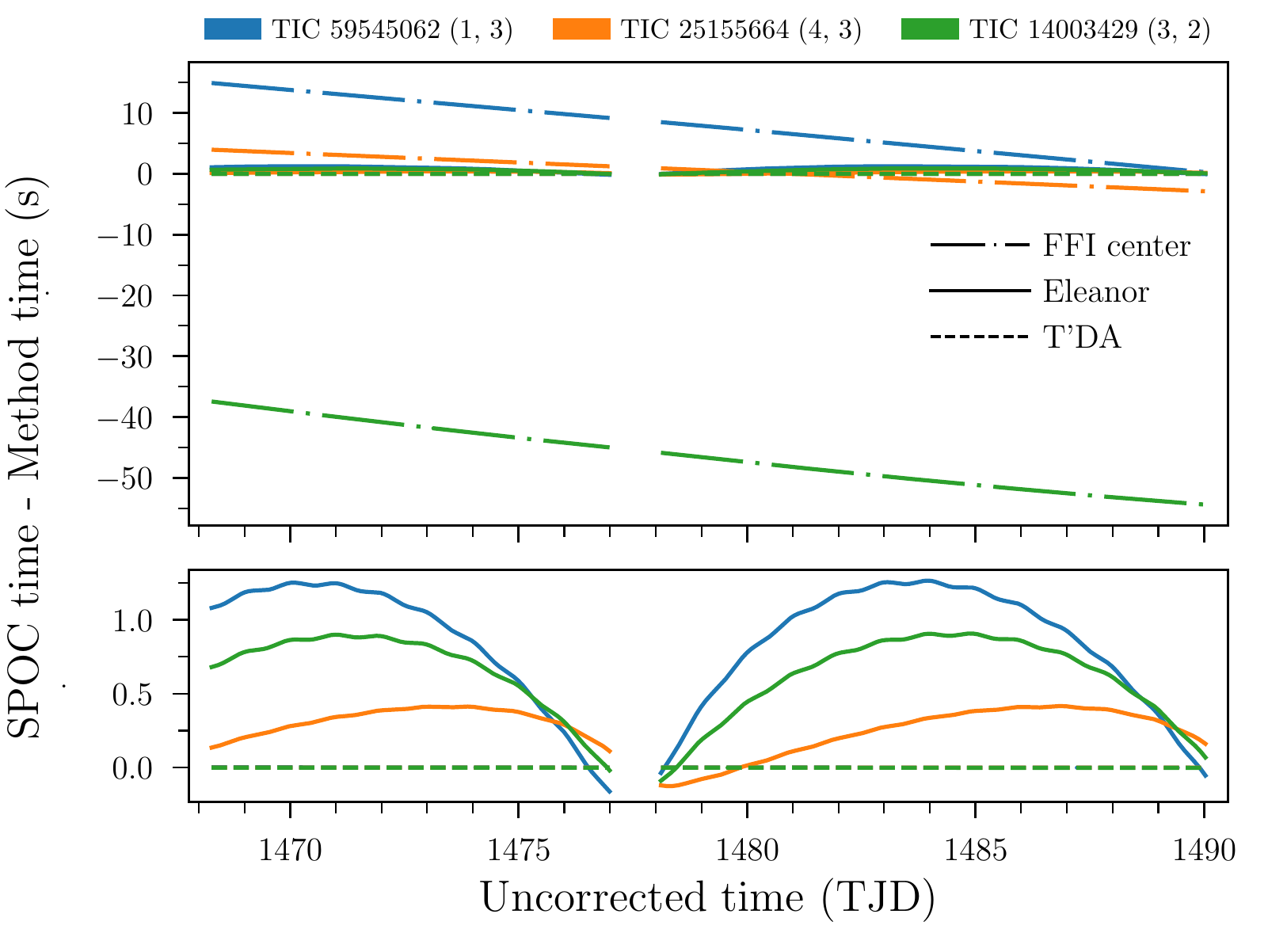}
    \caption{Comparison between timestamps provided in SPOC TPF data against times obtained from corresponding FFI data using different reference points for the time calculation. Differences are plotted against the uncorrected times. The comparison is shown for three targets observed in Sector 6 with different colors according to the top legend. The numbers in parentheses after the TIC ID indicate the position of the star in the FOV in terms of ``camera'' and ``CCD''. The targets were selected to cover a large portion of the FOV. Line styles indicate the method used for the calculation of times (see \sref{sec:timestamps}). The bottom panel provides a zoomed-in view of the y-range close to zero to better show the time differences obtained by the method adopted in this work and by the \code{Eleanor} pipeline \citep{eleanor}.}
    \label{fig:timecorr}
\end{figure}
We have compared the new timestamps generated in this way with the timestamps given in the TPFs provided by the SPOC. In all cases we arrive at timestamps within less than a millisecond from the ones in the TPFs. For TPFs we do not use this procedure, since these already have timestamps calculated for the target in question.

\fref{fig:timecorr} gives a comparison between the times computed using different reference points for three targets in Sector 6 at different locations in the TESS FOV against the times provided by SPOC in the TPF data. The ``FFI center'' method refers to the times provided in the FFI files, and used, \eg, by \code{TESSCut} \citep{tesscut} (which in turn is used internally by several other photometry pipelines), where the times can be off by several tens of seconds. Shown is also the reference used in \code{Eleanor} \citep{eleanor}, where the reference point is defined to be the Greenwich observatory -- this reduces the error on the times, but the resulting differences now have oscillations on timescales of both the TESS orbit and the Earth rotation (Greenwich is positioned on the Earth surface, hence periodically varying its distance to the observed stars). As mentioned above, the differences obtained in the TASOC pipeline from the SPICE kernels are consistently at the sub-millisecond level.

As can be seen in \fref{fig:timecorr}, the timing differences are generally within a few seconds for \code{Eleanor}, and can be on the order of minutes for \code{TESSCut}. Such offsets will likely have only have small effect on scientific analysis for most targets. However, in some cases, such as studies relying on the phase-stability for high-amplitude coherent oscillators or transit timing variations for exoplanets, the analysis could be affected by such offsets \citep{tess_time}, which in the case of \code{Eleanor} would also have periodic components besides the yearly periodicity. We note also that any potential impact will be increased with the reduction in the FFI cadence from \LC to \LCC in the extended mission.

From TASOC data release 5 and onward the timestamps for both FFIs and TPFs are corrected for the known constant time offsets \citep[see][]{tessdr25,tessdr27memo,tessdr29memo}, including earlier TESS data where this was not yet taken into account (before TESS data release 30). This includes effects from both ``staggered readout'' (FFIs) and detailed correction to start and end times (FFIs and TPFs).

\section{Very Bright Stars}\label{sec:bright_stars}
The brightest stars observed by TESS provide unique challenges. Stars brighter than a \Tmag of ${\sim}6$ saturate the detector, with excess charge spilling along the CCD columns. As these bleed columns become longer for very bright stars, larger target pixel files are required to capture the flux. Problems can be encountered when the bleed columns become excessively long.

The most significant problem arises when the bleed column extends beyond the region covered by the \ac{TPF}. This can arise either because the stamp selected for the target was too small, or because the star is too close to the edge of the CCD. In these cases the light curves cannot be constructed by simple aperture photometry. However, they may still be constructed from the `halo' photometry of scattered light and the wings of the point spread function (PSF) (\sref{sec:halo}).

For particularly bright stars near the edge of the CCD, the saturation can cause calibration problems. The smear correction, intended to remove the photometric smear that occurs during the shutterless readout, is calculated as the robust mean across virtual rows 2059--2068. If these pixels become saturated, as has happened with, \eg, \object{Procyon} and \object{$\alpha$~Cen}, the smear can be significantly overestimated along these columns. To rectify this, the \ac{TPF} needs to be recalibrated with a better smear correction.
The recalibration is done starting from the raw counts. In addition to the smear correction, it also differs from the pixel-level calibration applied in the SPOC pipeline by including an undershoot correction. All other calibration steps are peformed as normal \citep[See][]{TESSInstrumentHandbook,KDPH-pixel}.

The new smear correction is estimated from the science pixels themselves, selecting background pixels that do not contain light from any stars, and calculating a robust mean for each column from these pixels. Because this includes both smear and background flux, we estimate the background from the difference between this and the standard smear correction for columns unaffected by saturation, and remove it from our new smear correction. This new smear correction can then be applied.

Undershoot is a signal distortion that appears as a slight depression of the signal in a pixel that is read out after a bright pixel. This is generally a small correction for TESS, and is subsequently not applied in the SPOC pipeline \citep[see][]{TESSInstrumentHandbook}. However because it is most significant for light-to-dark transitions as occur around bright stars we have included the correction in our recalibration.

\begin{figure}
    \centering
    \includegraphics[width=\columnwidth,clip]{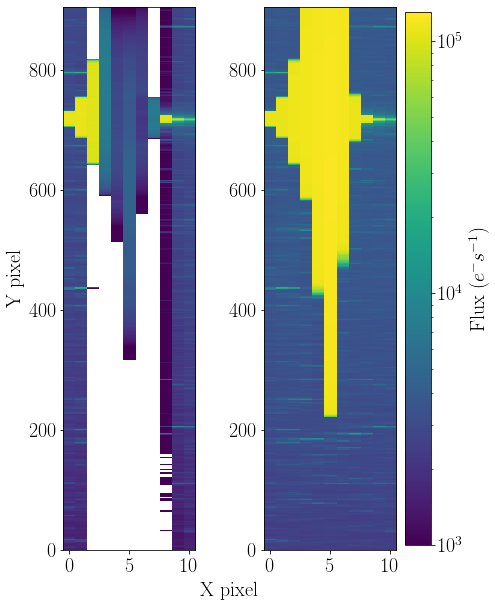}
    \caption{$\alpha$~Centauri ($\Tmag\!=\!-0.1$) as observed by \TESS in Sector 11. Left: Original flux with improper smear-column correction. Right: Recovered flux after re-calibration.}
    \label{fig:smear-recorrection}
\end{figure}

\fref{fig:smear-recorrection} shows a single frame from the Sector~11 \ac{TPF} of \object{$\alpha$~Cen} before and after this recalibration procedure has been applied. For stars such as \object{$\alpha$~Cen}, or any star that happens to share a column with these stars, this procedure makes previously unusable data viable. To compare how well this  procedure performs relative to the standard calibration of the SPOC pipeline, we can apply it to a bright star that is not affected by an overestimated smear correction, such as \object{$\beta$~Hyi} in Sector~1, and see how the flux of each pixel at each cadence differs. Despite the smear and undershoot calibration differences, the median ratio of the recalibrated flux to the SPOC flux for all pixels in the Sector 1 \ac{TPF} of \object{$\beta$~Hyi} is 0.999, with a 16th percentile of 0.986 and an 84th percentile of 1.006.

An additional issue for bright stars is that in each Sector the brightest stars (typically with $\Tmag\lesssim 2$) with large pixel stamps are not processed by the photometric module of the SPOC pipeline. As a result the \acp{TPF} provided by SPOC are without flux uncertainties, background fluxes, or an aperture calculated or even basic pixel calibrations (Sectors 1 and 2 only). Because of this, these targets are currently not processed by the TASOC photometry pipeline either. We are working on the implementation of basic pixel-level calibrations on these targets, enabling high-quality photometry for these as well.

Over the large photometric apertures used for bright stars, the background calculated from the SPOC photometric pipeline can be a poor match for the spatially-varying background, because it is uniform across the image. This can have the effect of introducing spurious signals, which can be especially severe in the case of halo photometry (\sref{sec:halo}), where the fainter parts of the PSF can be of comparable flux density to the background \citep{eisner19}.
We are also considering if the background calculated from the SPOC photometric pipeline is appropriate for less bright stars in terms of capturing spatial and temporal variations. Therefore, we have experimented with allowing for more spatial structure in the background used for TPF files, which is also allowed to vary in time. The strategy is to utilize the full background-structure estimated from the FFIs (\sref{subsec:background_estimation}),  interpolating it onto the higher temporal cadence, and then scale it to have the same mean as estimated from the background-pixels in the TPF. Thereby we can utilize the structural information from the FFIs, combined with the temporal resolution from the TPFs.

We found that this helped partially remove spurious signals from scattered light in some cases, but was not sufficient in other cases. To improve this we also added the possibility of additionally fitting a low-order 2D polynomial light map on top of the FFI structure, which of course comes at the risk of over-fitting. Additional work is still needed to find the optimal strategy and this feature is not yet enabled for the current data releases.

\section{Photometric Methods}\label{sec:photometry}
The central module in the pipeline is the ``Photometry'' module, which is responsible for the extraction of light curves from the FFIs and TPFs, after the initial preparations and restructuring of data.
The pipeline is built in a modular fashion, so different methods for photometry can be implemented into the same framework without duplicating functionality that is of a more general concern to CCD photometry.
This is implemented by having a general class, \code{BasePhotometry}, which defines the general concept of ``a photometry of a single target'', that implements all common functionality between the photometric methods, described in the following sections.
All specific photometric methods inherits all methods and properties from the \code{BasePhotometry} class, thus defining an abstraction layer between the data and the photometric methods. In this way, the individual methods do not need to worry about low-level data input/output, but can focus on the task of extracting the brightness of the target star as a function of time from the provided postage-stamp images provided by this underlying system.

When the photometry of a given target is started, the underlying \code{BasePhotometry} class will automatically create a ``postage-stamp'' cutout of the FFI or TPF in question, centered on the target, and load the necessary data into memory. The size of the stamp is scaled by the \Tmag of the target, creating larger stamps for bright targets (potentially thousands of pixels, as can be seen in \fref{fig:smear-recorrection}), and scaling down to a minimum stamp size of $11\!\times\!11$ pixels for fainter targets (see \fref{fig:stamp}). 

\code{BasePhotometry} provides functionality for accessing the postage-stamp images, uncertainty images, extracted background images and average image in several ways, always returning only the pixels within the postage-stamp, \ie, without having to load entire \ac{FFI} into memory. It also defines functionality for changing the size of the postage stamp, should any photometric method require it.

An interface for obtaining lists of all stars that are present within the current postage-stamp is also available.
This is created using the WCS solution to map the footprint of the postage-stamp onto the sky and querying the catalog files (see \sref{sec:catalogs_and_todo}) for all targets that fall within these coordinates (plus-minus a small buffer) at the time of observation. The resulting list contains both the TIC identifiers, sky- and pixel-coordinates and the \Tmag of the targets. Furthermore, it can provide the same catalog of targets where the pixel-coordinates take into account the spacecraft pointing jitter, which depends on the time and position on the FOV, meaning the catalog is time-dependent. We will return to this in \sref{sec:psf}.

When starting the photometric reduction, \code{BasePhotometry} will also define a light curve table, which will contain the timestamps (see \sref{sec:timestamps}), quality flags, cadence numbers and any other relevant information corresponding to each postage-stamp image as a function of time. Columns for measured flux and corresponding uncertainly are also created, but initially left empty. It is then up to the specific photometric algorithms to populate these columns of the table.

Once the photometry has populated the light curve table with measured flux and uncertainties, \code{BasePhotometry} also provides the functionality for saving the extracted light curves to FITS files, with all the appropriate headers and meta-information. The format of the output files will be described in \sref{sec:data_formats}. At the same time, a set of diagnostic information about the extracted light curve is calculated and stored into the TODO-file for use in the following steps of the pipeline. This includes, for instance, the median flux level, RMS- and point-to-point noise and a variability metric (see Lund \etal, in prep., for details).

\subsection{Aperture Photometry}\label{sec:aperture}
The primary method for extracting light curves in the pipeline is simple aperture photometry, where the total flux of the target is summed up within a pixel mask containing, ideally, only the target in question. The approach taken builds heavily on the K2P$^2$ \citep{K2P2,Handberg2017} method, which was developed for the K2 mission, but with a few modifications.

The method define apertures by means of \code{DBSCAN} \citep[Density-Based Spatial Clustering of Applications with Noise;][]{dbscan}, as implemented in \code{scikit-learn} \citep{scikit-learn}. This is applied to pixels with a flux level above a given threshold, given by the mode plus $0.8$ times the standardized MAD of the flux in the pixel stamp. A constraint of a minimum of 4 pixels is placed on the size of the clusters such that, \eg, single high-flux pixels are discarded as being noise. In the case that several distinct clusters are identified the position of the target in question is used to identify to which pixel cluster the target belongs. 

\begin{figure*}
    \centering
    \includegraphics[width=\textwidth]{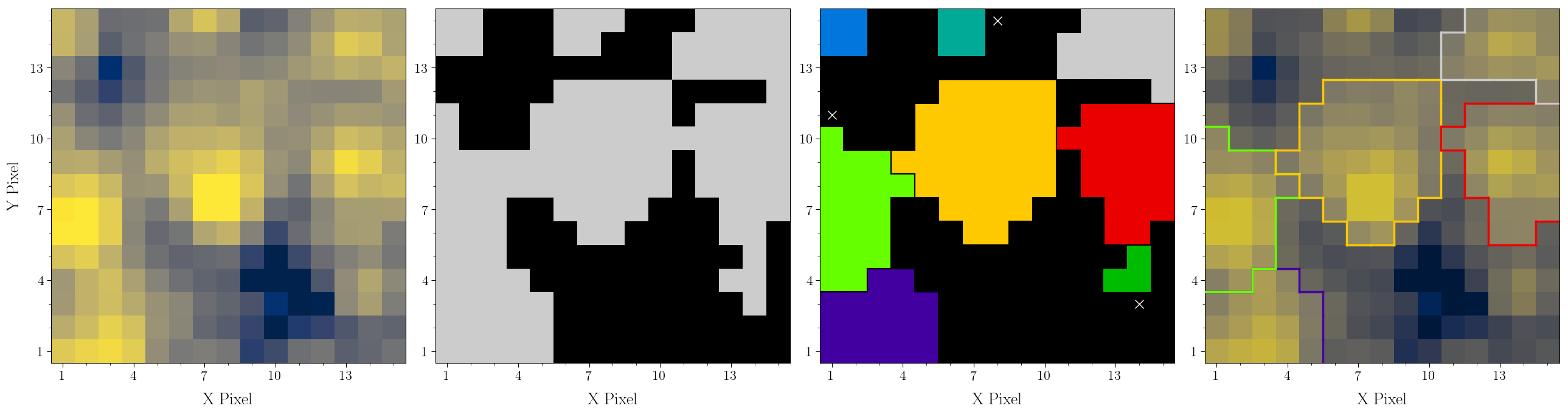}
    \caption{Example of apertures set using the described procedures for a target in a relatively crowded region, here for the star TIC~410446575 in Sector 3 (central yellow aperture). Going from the left, the first panel shows the summed image of the pixel cutout used for the star; the second panel gives in grey the pixels with flux levels above the adopted threshold; the third panel shows the apertures found from applying the clustering and watershed algorithms, pixels identified as noise are marked with a white ``x''; the fourth panel shows the outlines of the final aperture for the cutout after discarding apertures with four pixels of less.}
    \label{fig:apertures}
\end{figure*}

After the identification of pixel clusters the watershed image segmentation from \citet{scikit-image} is run to split clusters that may contain two or more close targets. In a departure from the approach of \citet{K2P2} we ensure that all basins used for the watershed, which are found from a local peak-finding algorithm, can be associated with a target from the TIC that lies within the pixel stamp. In the end, the set of pixels from the clustering and watershed segmentation that contain the position of the target star is returned as the aperture to use for the photometric extraction.
If no pixels are found above the flux threshold, or if the known position of the target star does not lie within any of the identified sets of pixels a minimum $2{\times}2$-pixel aperture is returned at the target position and a warning flag is set (See \tref{tab:bit2}). 
\fref{fig:apertures} provides two examples of the definition of apertures, one for an isolated star and one for a star in a crowded field. As seen, the clustering combined with the image segmentation allows apertures to be defined even in relatively crowded regions.
We note that for \SC data, where the adopted pixel stamp is taken from the \TESS mission, we also perform photometric extraction for potential secondary targets found in the stamp. So, while we currently only include primary targets with $\Tmag<15$ in the processing, some secondary targets in the \SC stamps will have magnitudes beyond this limit. These additional \SC targets are most clearly seen as the sparse points in Figures~\ref{fig:rms}--\ref{fig:magtoflux} beyond $\Tmag = 15$.

When the aperture belonging to the target star has been identified it is checked if any of the aperture pixels lie at the edge of the pixel stamp (only for \LC data). If this is the case the stamp is extended by 5 pixels in the direction of the edge containing aperture pixels and the aperture definition is run anew. The initial size of the stamp is determined by the TESS magnitude of the target star, with a relation optimized to return the smallest stamp without the need for resizing (\fref{fig:stamp}). So, while the resizing is allowed to take place 5 times this rarely happens. If the resizing limit is reached an error flag is set for the star. 

When the final aperture has been defined a contamination metric is computed if the catalog indicates that stars other than the target star lie within the aperture. The contamination metric, $C$, is given by one minus the flux ratio of the target star to all stars in the aperture, including the target. In practice it is calculated from the sum of catalog magnitudes of stars located within the defined aperture:
\begin{equation}\label{eqn:contamination}
    C = 1 - \frac{f_\mathrm{target}}{f_\mathrm{total}} = 1 - 10^{0.4(\Tmag_\mathrm{total} - \Tmag_\mathrm{target})}
\end{equation}

In this manner the contamination will tend to 1 if the flux from secondary stars equal that of the target star. See \sref{sec:validate} and \fref{fig:con} for the cuts made in the data validation based on the contamination metric. We note that the metric will not include flux from stars just outside the aperture, so the metric can be seen as a minimum value for the flux contamination.
We do not currently attempt to correct the flux level of the target star for crowding effects.

\subsection{Halo Photometry}\label{sec:halo}
For the very brightest stars, saturation produces a bleed column that is so long that it reaches the edge of the CCD chip, or which is otherwise too saturated to include in a photometric aperture. For these stars we employ the halo photometry method \citep[\code{halophot}:][]{halo,halopope}, which constructs a calibrated light curve from unsaturated pixels in the deep PSF (the `halo').

The light curve is constructed as a linear combination of the time series of the individual unsaturated pixels in the vicinity of the bright star. The weights are determined by gradient-descent optimization to minimize an objective function of the resulting light curve, the generalized `lagged Total Variation' \citep{halopope}. This is a convex objective function (`TV-min'), and has analytic derivatives which can be computed with automatic differentiation, for which we use \code{autograd} \citep{autograd}. 

\code{halophot} has been shown on K2 data to produce light curves with comparable noise to conventionally-observed unsaturated stars, without significant loss of sensitivity to planets or stellar variability \citep{halopope}. 

In our TESS implementation, we first adopt a default circular aperture with a 20-pixel radius centred on the position of the star. We then discard saturated pixels by an iterative process: using a conservative range of estimates of the number of saturated pixels, we apply the TV-min algorithm to pixel sets with one more pixel discarded at a time. Pixels that are saturated contribute very little Total Variation, and therefore when we have cut out all saturated pixels there is a significant increase in the standard deviation of the resulting light curve. Next, background stars are identified using \code{DBSCAN}, similarly as done in aperture photometry (\sref{sec:aperture}), and masked out. We then apply the TV-min algorithm to this censored array of pixels. The resulting light curve is scaled to absolute flux using the relation described in \sref{sec:validate}. The resulting weight-maps, indicating the final weights given to each pixel, are stored as an extra extension in the final output FITS files (see \sref{sec:data_formats}).

\subsection{PSF Photometry}\label{sec:psf}
The TESS Point-Spread Functions (PSFs) are provided by the \TESS team and give the two-dimensional sub-pixel distribution of light from a point source as measured by the TESS spacecraft. The PSFs were created in-flight by micro-dithered observations of bright isolated stars, resulting in PSFs measured at 25 positions spanning each CCD \citep[see][]{TESSInstrumentHandbook}.
These PSFs can be linearly interpolated to any position in the FOV, and integrated onto the pixel-grid to yield the Pixel Response Function (PRF), which is the ``pixelated'' version of the PSF.

The objective of PSF Photometry is to fit this inferred shape of a star to the measured images, assuming a good catalog of stars in the image. In the standard PSF photometry, this means we have three free parameters per star in the image: position in pixels and flux. A nonlinear optimization algorithm is used to optimize all the free parameters by, at each iteration, re-integrate the underlying PSF to a new PRF for each star and match this to the observed image.

However, the TIC (see~\sref{sec:catalogs_and_todo}) contains the positions of stars with very small uncertainties, especially when compared to the 21\arcsec\ pixels of \TESS. It is therefore a reasonable assumption that we can fix the positions of the stars for a given image using the TIC. In that case the PSF fitting problem becomes linear, and can be written in the following way:
\begin{equation}
    \begin{bmatrix} \vdots &  & \vdots \\ \mathrm{PRF}_{i,1} & \cdots & \mathrm{PRF}_{i,N} \\ \vdots & & \vdots \end{bmatrix} \begin{bmatrix} \vdots \\ f_j \\ \vdots \end{bmatrix} = \begin{bmatrix} \vdots \\ p_i \\ \vdots \end{bmatrix} \, ,
\end{equation}
where $f_j$ is the desired flux of the $j$th star, $p_i$ is the measured flux in pixel $i$ and $\mathrm{PRF}_{i,j}$ is the PRF of the $j$th star in pixel $i$. This problem can be solved much faster, using standard linear algebra routines, which significantly speeds up the calculation.

As a final step, the flux is adjusted by performing aperture photometry on the flux in the residual image in a small aperture ($2\!\times\!2$ pixels) centered on the target star, and adding this to the extracted flux. This is inspired by \citet{MOMF} and will correct the measured flux for small deviations near the PSF center.
Preliminary tests show that this correction can be on the order $\sim$1\% in the case of a $\Tmag\!=\!7$ star, which is near the saturation limit of the detector.

Both the standard PSF fitting and the linear PSF fitting are implemented in the pipeline, but are currently not used. One shortcoming of PSF photometry is its inability to handle bright saturated stars, which often are the ones of most interest to asteroseismology. This is due to the PSF model not being a valid representation of the observed image due to saturated pixels and bleed-trails causing deviations unique to the brightness and position of the star on the detector. In future data releases, PSF photometry could be used for non-saturated stars in crowded regions, where aperture photometry becomes inefficient and prone to errors from contamination. 

\subsection{Difference Imaging}
The last method of photometry currently implemented in the TASOC Photometry pipeline is difference imaging, where aperture photometry is performed on the difference between the image in questions and a reference-image. The method is well established and has been demonstrated to work on \TESS data in crowded fields \citep[See][]{cdips}. The implementation in the pipeline is currently not at a stage where it can be run in a routine manner and is therefore currently not in use. We are, however, planning on including this in future releases for use in crowded fields.

\section{Parallel processing}
\begin{figure}
    \centering
    \includegraphics[width=0.8\columnwidth]{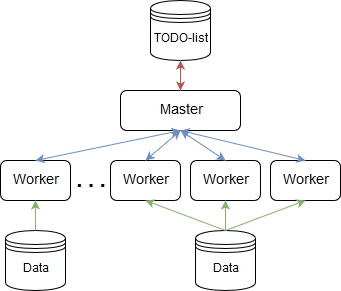}
    \caption{Schematic overview of the computational setup used for the majority of the TASOC Pipeline. The Master process keeps track of the tasks to be processed using the TODO-list and communicates the tasks to the workers. Each worker has direct access to the input data through a shared or distributed file system.}
    \label{fig:processing}
\end{figure}

In most of the TASOC pipeline setup, the same basic setup for processing the many tasks in a parallel scalable way is used. The setup is illustrated in \fref{fig:processing} and is also used in the corrections and classifications parts of the full TASOC pipeline. This consists of a master process which hands out tasks to workers, which upon completion return the results of that task back to the master. This is an efficient way to perform a large number of independent tasks when there are more tasks than workers, especially when the run times can vary for each task.
In practice this is done by utilizing a combination of Python's \code{multiprocessing} module, Message Passing Interface (MPI) and the Slurm workload manager\footnote{\url{https://slurm.schedmd.com}}.

Specifically for the photometry part of the pipeline, as discussed in this paper, splitting the work into independent sub-tasks is done in two different ways:

The parallelization is fairly trivial in the initial ``prepare'' step, since each FFI can be treated independently in the estimation of backgrounds and pixel-level flags (\sref{subsec:background_estimation}). Each worker can therefore handle all the loading of the FFI and all calculations independently as a single ``task'' and the master process only has to receive the output of each worker, perform any time-dependent calculations, and save the results to disk.

For the photometry itself, we define a ``task'' as being the photometry of a single star. The master is responsible for finding the next task to be processed from the TODO-list (see \sref{sec:catalogs_and_todo}) and passing this to an idle worker. The worker, upon receiving a task, will perform the required photometry (see \sref{sec:photometry}), save the resulting light curve to file (see \sref{sec:data_formats}), and return the status and additional diagnostic information from the photometry to the master process. Based on the returned information from the worker, together with the information in the TODO-list, the master can make decisions on which task to process next, if any other tasks needs to be skipped, or if additional processes needs to be added.
An example where this is used is if fainter stars are found to fall within the derived aperture of a brighter star. The master will mark this in the TODO-list, so photometry is not subsequently attempted on these fainter stars, since that would result in the same aperture as found for the bright star. Complete records of which stars were skipped, and which brighter stars responsible, are also stored in the TODO-list.

Using this overall computational setup, we are able to use the power of a distributed computing systems, while still being able to dynamically handle dependencies between tasks, like the ones described above. Currently this capability is only moderately exploited, but could be used to dynamically change the tasks to be performed on-the-fly based on the information extracted up until that point, \eg, changing the type of photometry based on the crowding of a particular region of the FOV.

\section{Light curve Data Formats}\label{sec:data_formats}
The primary data format for extracted and corrected light curves is FITS (Flexible Image Transport System\footnote{\url{https://fits.gsfc.nasa.gov/fits_standard.html}}), and is provided in a compressed Gzip format.
A FITS light curve file produced by T'DA and stored on TASOC will be named following the structure:
\begin{center}\footnotesize
    \code{tess$\{$TIC ID$\}$-s$\{$sector$\}$-$\{$camera$\}$-$\{$ccd$\}$-c$\{$cadence$\}$\\-dr$\{$data~release$\}$-v$\{$version$\}$-tasoc$\_$lc.fits.gz}
\end{center}
The ``TIC ID'' (TESS Input Catalog identifier) of the star is zero (pre-)padded to 11 digits, the ``sector'' is zero pre-padded to 3 digits, the ``cadence'' is in seconds and zero pre-padded to 4 digits, the ``data release'' is zero pre-padded to 2 digits and refers to the official release of the data from the mission, the ``version'' is zero pre-padded to 2 digits and refers to the TASOC data release (counting from 1). As an example, the star TIC~62483237, observed in Sector 1 on camera 1, CCD 2, in \SC cadence and part of the first data release and fifth TASOC processing will have the name:
\begin{center}\footnotesize
    \code{tess00062483237-s001-1-2-c0120-dr01-v05-tasoc$\_$lc.fits.gz}
\end{center}

Note that files released prior to TASOC data release 5 did not include the camera and CCD in the file names.

Each light curve FITS file has four extensions: a ``\code{PRIMARY}'' header with general information on the star and the observations; a ``\code{LIGHTCURVE}'' table with time, raw flux, corrected flux, etc.; a ``\code{SUMIMAGE}'' with an image given by the time-averaged pixel data; and an ``\code{APERTURE}'' image. The information provided in the FITS file is intended to mimic that provided in the official \TESS products -- please consult the ``TESS Science Data Products Description''\footnote{\url{https://archive.stsci.edu/missions/tess/doc/EXP-TESS-ARC-ICD-TM-0014.pdf}} for more information.

Note, targets processed with the Halo photometry option (see the \code{PHOTMET} key in primary FITS header for the adopted photometry method) have the additional extension ``\code{WEIGHTMAP}'' in their FITS file, which gives the weight assigned to each pixel in the Halo photometry method.

From file version 1.3 additional columns have been added to the ``\code{LIGHTCURVE}'' table containing quality flags.
One of these, ``\code{PIXEL$\_$QUALITY}'', contains the quality flag provided by the \TESS team. For an explanation to the bit values used here see the  \href{https://outerspace.stsci.edu/display/TESS/2.0+-+Data+Product+Overview}{TESS Archive Manual}.
The column ``\code{QUALITY}'' gives the quality flags set by the TASOC pipeline, which have the meanings as shown in \tref{tab:bit}.

With file version 1.4 a few additional keywords have been added to the ``\code{LIGHTCURVE}'' header. Of particular notice is the ``\code{XPOSURE}'' key, which gives the actual exposure of the observations, taking into account dead-time from readout and from the cosmic ray mitigation \citep[see][]{CosmicRayMitigation,TESSInstrumentHandbook}. Using this value for the integration of measured flux will, for instance, be important for the calculation of signal apodization.

\begin{table*}[h!]
    \footnotesize
    \centering
    \begin{threeparttable}
    \caption{TASOC ``\code{QUALITY}'' flags.}
    \begin{tabular}{@{}rrp{100mm}@{}}
    \toprule
    Bit & Value & Description\\
    \midrule
    0 & 0 & All is OK.\\
    1 & 1 & Data point flagged as bad based on quality flag by \TESS team (their bits 1, 2, 4, 8, 32, 64, 128, and/or 4096).\\
    2 & 2 & Manually excluded by TASOC team.\\
    3 & 4 & Data point has been sigma-clipped.\\
    4 & 8 & A additive constant jump correction has been applied.\\
    5 & 16 & A additive linear jump correction has been applied.\\
    6 & 32 & A multiplicative constant jump correction has been applied.\\
    7 & 64 & A multiplicative linear jump correction has been applied.\\
    8 & 128 & Data point has been interpolated.\\
    9 & 256 & Data point has been rejected in processing.\\
    \bottomrule
    \end{tabular}%
    \label{tab:bit}
    \end{threeparttable}
\end{table*}

With file version 5 onwards a photometric data validation (\code{DATAVAL}) flag has also been added to the ``\code{PRIMARY}'' header. These flags have the following meanings:

\begin{table*}[h!]
    \footnotesize
    \centering
    \begin{threeparttable}
    \caption{TASOC ``\code{DATAVAL}'' flags.}
    \begin{tabular}{@{}rrp{100mm}@{}}
    \toprule
    Bit & Value & Description\\\midrule
    0 & 0 & All is OK.\\
    1 & 1 & Not in use.\\
    2 & 2 & Star has lower flux than given by magnitude relation.\\
    3 & 4 & Not in use.\\
    4 & 8 & Not in use.\\
    5* & 16 & Star has minimum 2x2 mask.\\
    6* & 32 & Star has smaller mask than general relation.\\
    7* & 64 & Star has larger mask than general relation.\\
    8 & 128 & Not in use.\\
    9 & 256 & PTP-MDV lower than theoretical.\\
    10 & 512 & RMS lower than theoretical.\\
    11* & 1024 & Invalid Contamination.\\
    12 & 2048 & Contamination high.\\
    13* & 4096 & Invalid mean flux.\\
    14* & 8192 & Invalid Noise.\\
    \bottomrule
    \end{tabular}%
    \label{tab:bit2}
    \end{threeparttable}
\end{table*}

Bits marked with a ``*'' in \tref{tab:bit2} have been used internally to identify targets to hold back from being released -- these targets will be scrutinized further and may be made available with a subsequent release. Therefore, only bits not marked with ``*'' will actually appear in the released data. The boundaries used for the flags are given in Figures~\ref{fig:rms}--\ref{fig:magtoflux}. We note that a target may have an aperture of 4 pixels (\ie the minimum allowed aperture) without being flagged with bit 5, because this bit is only set when the aperture definition has failed in some manner and has defaulted to the minimum $2{\times}2$. 

Users should make sure to check the photometric data validation (\code{DATAVAL}) flag of any specific star under study, as well as the aperture and sum-images.

\section{Data Validation and Performance}\label{sec:validate}

Below we describe in turn the different metrics used in the validation of the extracted photometry before release and for our assessment of the pipeline performance. Information on how the data validation results are reported can be found in \sref{sec:data_formats} and \tref{tab:bit2}.

\begin{figure*}

    \centering
    \includegraphics[width=\textwidth]{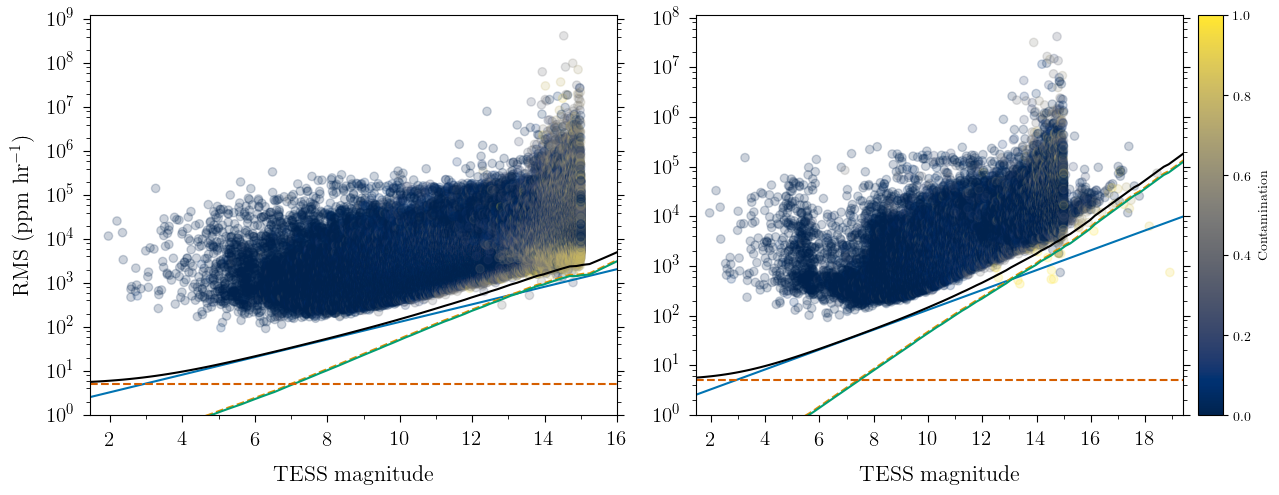}
    \caption{RMS noise on 1-hour time scale, here for stars observed in Sector 3. Left: targets extracted from \LC FFI data. Right: targets extracted from \SC TPF data. The lines give the  predicted noise estimates following \citet{Sullivan2015} (blue full: shot noise; green full: read noise; yellow dashed: zodiacal noise; red dashed: adopted systematic noise; black full: total noise).}
    \label{fig:rms}
\end{figure*}

\begin{figure*}
    \centering
    \includegraphics[width=\textwidth]{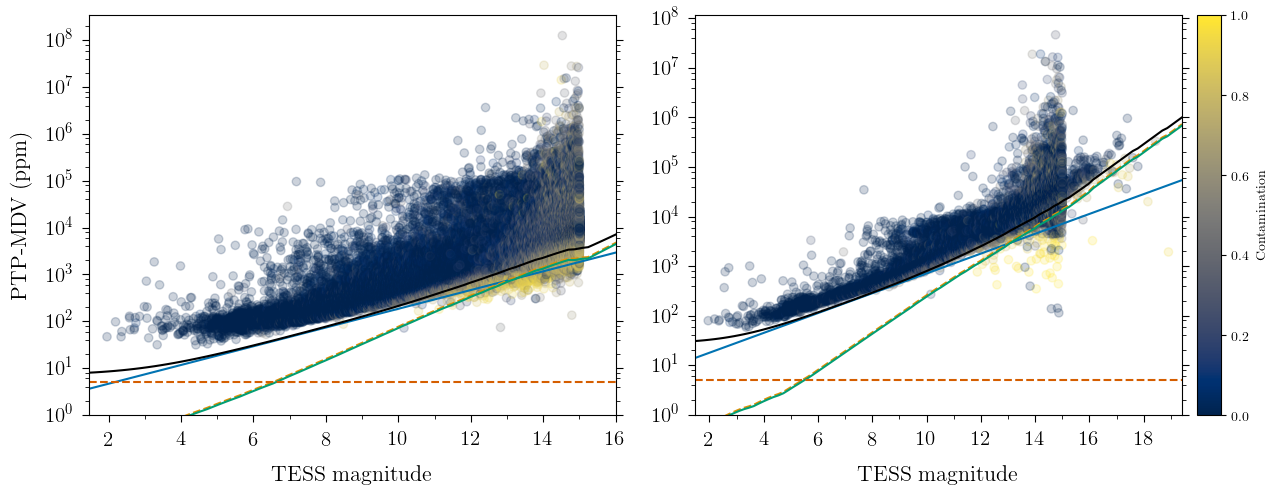}
    \caption{Point-to-point Median-Differential-Variability (MDV), here for stars observed in Sector 3. Left: targets extracted from \LC FFI data. Right: targets extracted from \SC TPF data. The lines give the  predicted noise estimates following \citet{Sullivan2015} (blue full: shot noise; green full: read noise; yellow dashed: zodiacal noise; red dashed: adopted systematic noise; black full: total noise).}
    \label{fig:ptp}
\end{figure*}

\begin{figure*}
    \centering
    \includegraphics[width=1\textwidth]{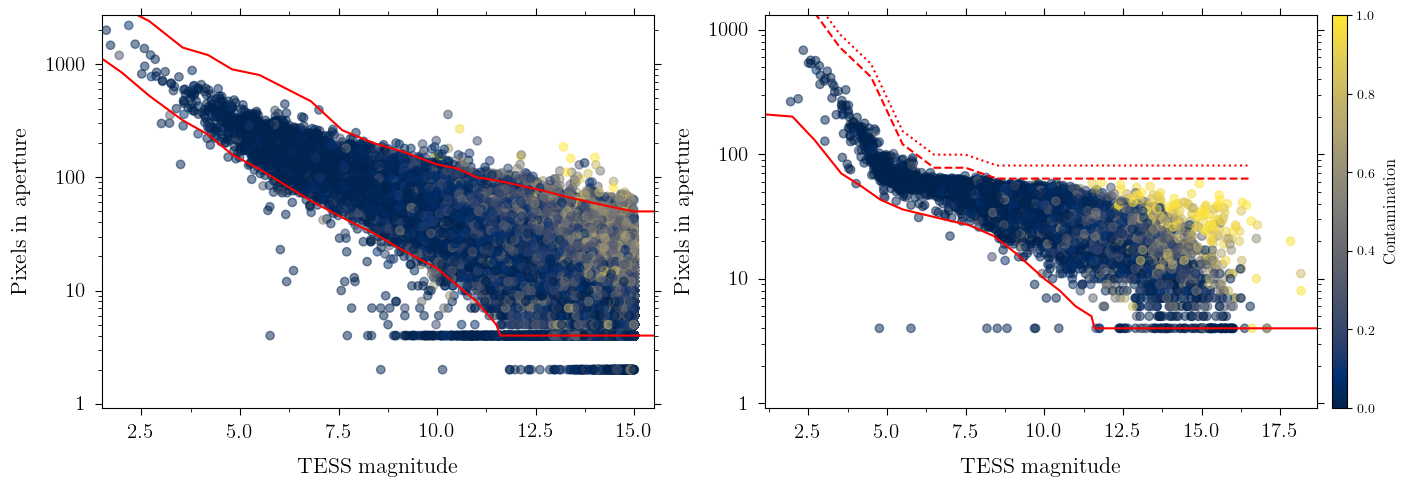}
    \caption{Pixel in apertures as a function of \TESS magnitude, here for Sector 1 data. The left panels show apertures for \LC targets, while the right panels show apertures for \SC target. The individual points are color-coded by the contamination. The full red lines give the boundaries for the data validation. The dotted line in the right panel gives median relation for the maximum allowed number of pixels given the stamp size, while the dashed line shows the corresponding relation under the assumption that the aperture is circular.}
    \label{fig:pix}
\end{figure*}

\begin{figure*}
    \centering
    \includegraphics[width=1\textwidth]{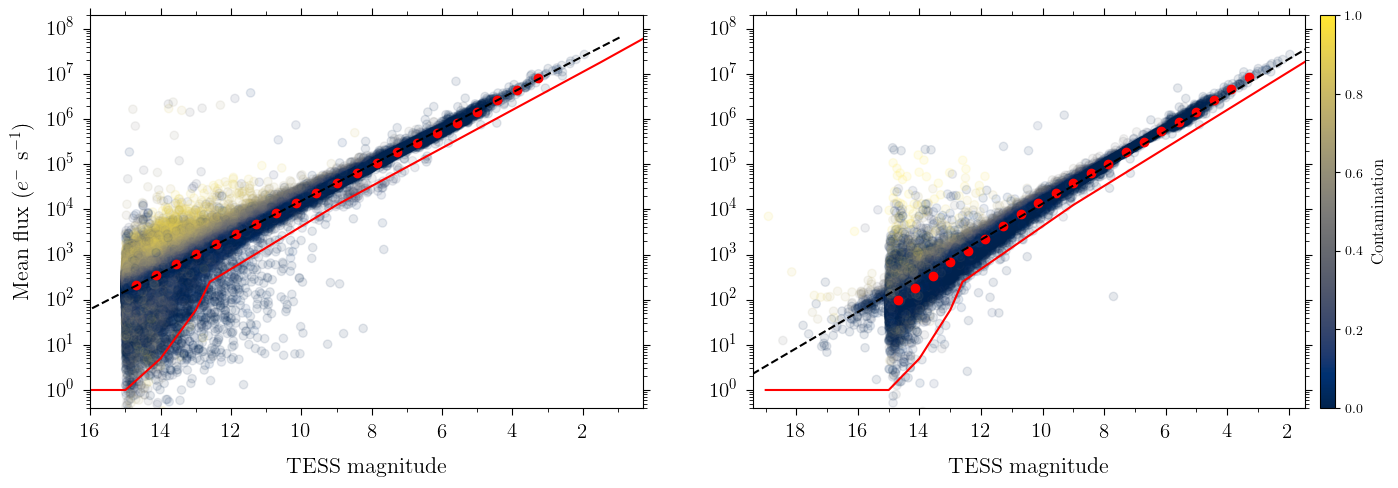}
    \caption{Relation between extracted flux from simple aperture photometry and the \TESS magnitude, color-coded by contamination, here for Sector 3 data. The left (right) panels show values for \LC (\SC) data. The black dashed line gives the relations obtained following the prescription in \eqref{eq:flux}, which has been fitted to the red median binned points. The full red line gives the adopted boundary for the data validation.}
    \label{fig:magtoflux}
\end{figure*}

\paragraph{Noise metrics}
The photometric quality of the reduced (raw) light curves is summarized in Figures~\ref{fig:rms}--\ref{fig:ptp}. \fref{fig:rms} shows the 1-hour root-mean-square (RMS) noise in parts-per-million (ppm) as a function of \Tmag; \fref{fig:ptp} gives the point-to-point Median-Differential-Variability (MDV) (corresponding to RMS on time scale of observing cadence). For the expected-noise curves \citep{Sullivan2015} we used relations for mean flux (\fref{fig:magtoflux}) and number of aperture pixels (\fref{fig:pix}) as a function of \TESS magnitude derived from the processed data. As seen the raw photometry generally follows the expected noise characteristics. A comparison of the noise metrics for systematics-corrected data will be provided in Lund \etal (in prep.).

\paragraph{Aperture size}
\fref{fig:pix} shows the sizes of the defined apertures as a function of \Tmag.
A minimum aperture of 4 pixels has been adopted for the TASOC processing -- targets with smaller apertures in \fref{fig:pix} are situated on CCD edges and have not been released (cf. \tref{tab:bit2}). 
The full red lines give the boundaries used in the data validation, defined by visual inspection of the pixel stamps at different magnitudes. For \SC cadence targets only a lower bound is used because the upper aperture limit is typically set by the downloaded stamp size. This restriction is clearly seen in \fref{fig:pix} (right panel) where the dotted line indicates the maximum number of pixels given the downloaded stamp if requiring that the aperture does not include the edge pixels of the stamp -- the dashed line gives the corresponding aperture size assuming the pixel stamp is square, and that the aperture is circular. As seen, this latter limit nicely follows the upper boundary of the obtained aperture sizes, implying that in the future the \TESS team might consider increasing the size of the stamp selected for \SC stars to reduce the risk of missing parts of the stellar flux. We caution also that one should be aware of contamination (see below), especially at high magnitudes. As seen from \fref{fig:pix} the faint targets with larger-than-average apertures are typically significantly contaminated.

\begin{figure}
    \centering
    \includegraphics[width=1\columnwidth]{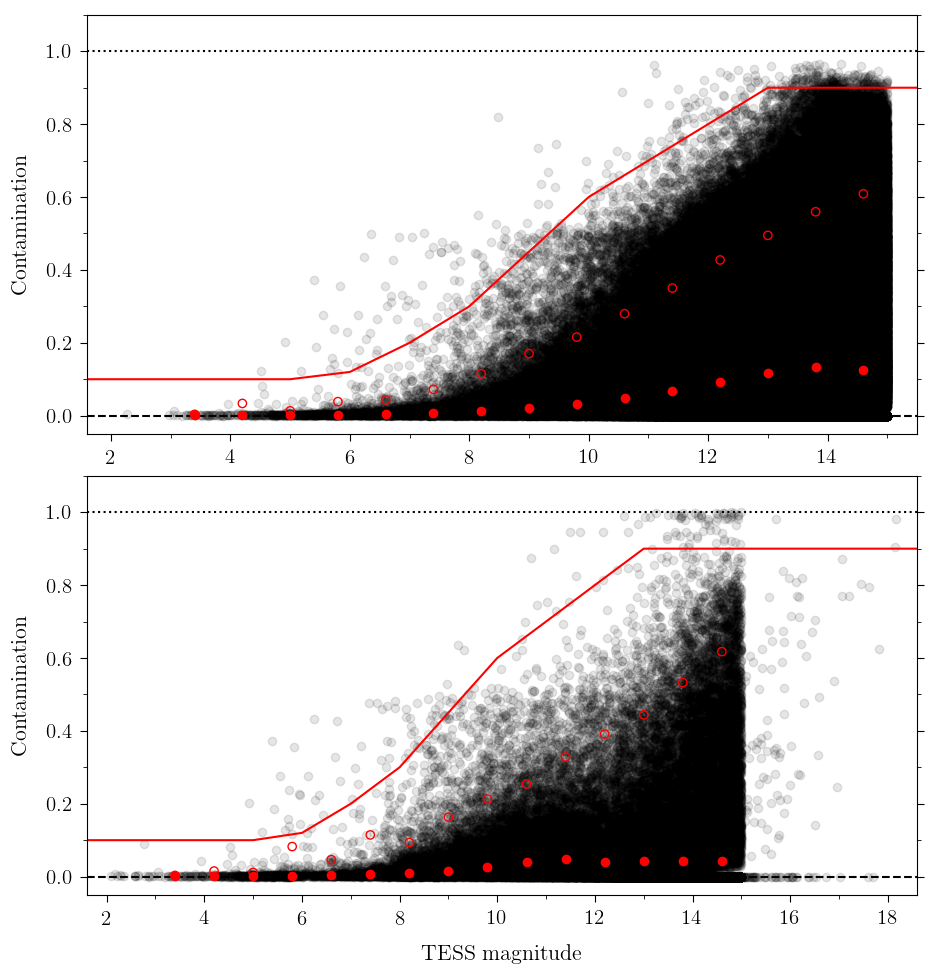}
    \caption{Contamination metric as a function of \Tmag, here for Sector 5 data. The top (bottom) panel gives contamination for \LC (\SC) data. The red full curve gives the boundary used in the photometry data validation. The red full points give the median binned values, while the empty red points give the binned $95$th percentile.}
    \label{fig:con}
\end{figure}

\paragraph{Contamination}
\fref{fig:con} shows the contamination metric for each star as a function of \Tmag (\sref{sec:aperture}; provided in the FITS light curve header as \code{AP$\_$CONT}). We encourage users of the data to keep this value in mind when interpreting signals extracted for a given star -- the metric gives the fraction of flux in the light curve contributed from stars other than the main one (see~\eqref{eqn:contamination}). Note therefore that flux in the aperture from a neighboring star that does not lie within the aperture is not taken into account. The \ac{WCS} provided with the ``\code{APERTURE}'' extension in the FITS file can be used to identify which other stars fall within the aperture of the main star.

\paragraph{Flux relation}
\fref{fig:magtoflux} shows the relation between the extracted mean flux for a star and its \TESS magnitude. This relation can be described well by the relation:
\begin{equation}\label{eq:flux}
    \Tmag \approx {zp} - 2.5 \log_{10}(S)\, ,
\end{equation}
where $S$ gives the mean flux in $e^{-}\,\, \rm s^{-1}$, and ${zp}$ is the zero-point for the relation. 
The fit of \eqref{eq:flux} is performed to median-binned values for a given Sector and cadence (see \fref{fig:magtoflux}), and considers only targets with a contamination below $0.15$. As expected, targets with a flux level above the median relation are typically found to have a significant contamination.
From Sectors 1-6 and considering both relations for \LC and \SC data, we find ${zp}$ values in the range 20.401--20.506. In general we find slightly lower ${zp}$ for \SC data as compared to \LC data, which likely can be attributed to the difference in the background correction, and possibly the fact that the \SC aperture size is limited by the downloaded pixel stamp (see ``aperture size'' above). It is also expected that the ${zp}$ will vary slightly from Sector to Sector due to differences in the overall crowding, and hence the relative importance of the sky background.
\eqref{eq:flux}, adopting a mean value of ${zp}=20.451$, is used for stars with photometry extracted using the Halo method (\sref{sec:halo}), in order to obtain the correct relative amplitudes. 

\begin{figure*}
    \centering
    \includegraphics[width=1\textwidth]{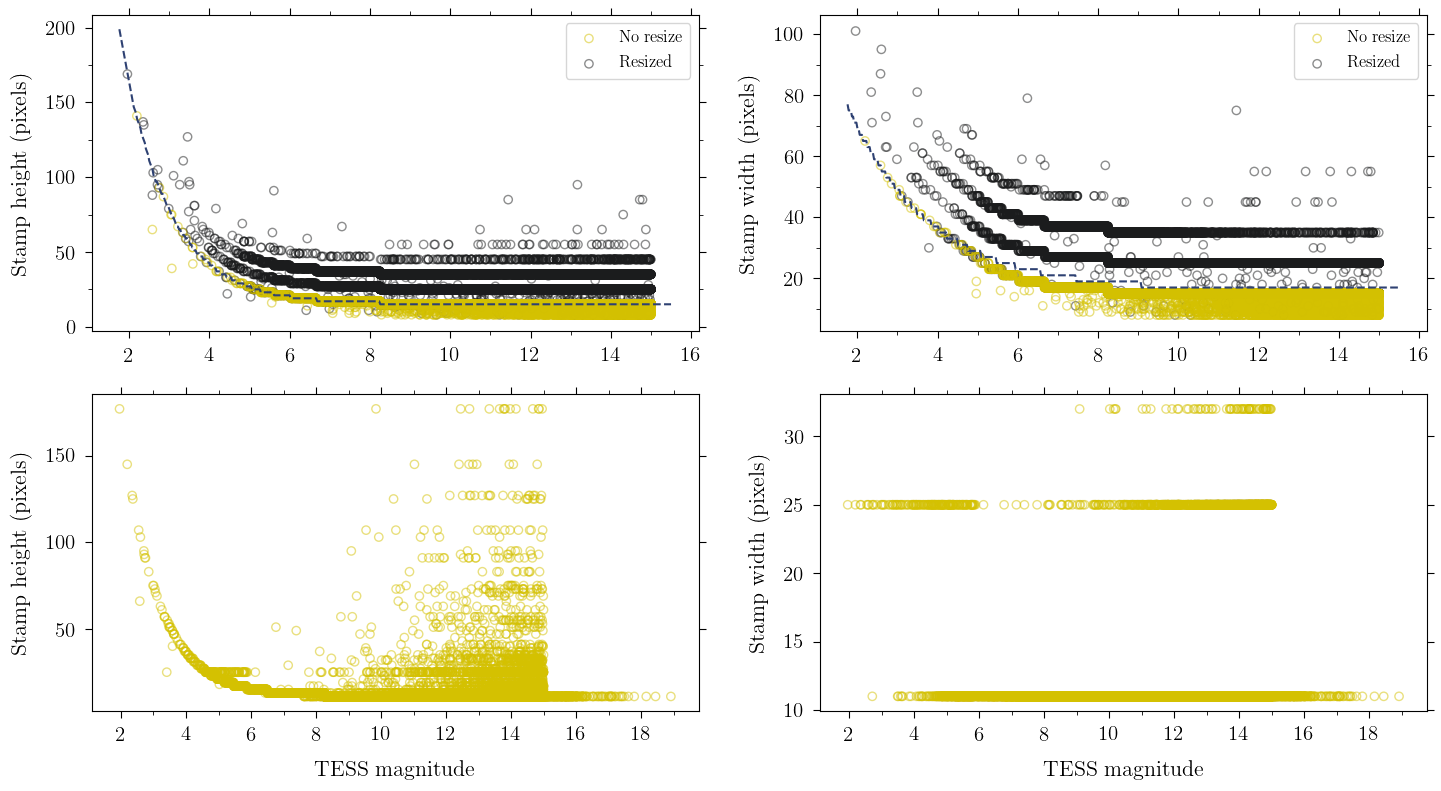}
    \caption{Relation between stamp height (left) and width (right) as a function of \Tmag, here for Sector 2 data. The top (bottom) panels show the values for \LC (\SC) data. Yellow points indicate stamps that have not been re-sized, while black points show values for re-sized stamps. The dashed line in the top panels shows the starting values of the stamp size.}
    \label{fig:stamp}
\end{figure*}

\paragraph{Pixel stamp size}
\fref{fig:stamp} shows the stamp sizes for the cut-outs made around each processed target. For \SC data the stamp provided by the \TESS mission is always used. In cases where a defined aperture touches the edge of the pixel stamp (for \LC data), the stamp is allowed to re-size in one or both directions by a fixed step of 5 pixels, and the aperture is defined anew (see also \sref{sec:aperture}). The starting guess for the stamp size (width and height) has been optimized to reduce the number of required re-sizes and thereby also processing time. The maximum number of allowed re-sizes is currently set to 5, at which state the photometry would stop, but this is almost never reached in practice. FFI targets seen to have heights/widths falling below the well-defined relation are found on the edges of the CCDs and, hence have a limited possible height/width. For secondary targets identified in TPF stamps a fixed aperture of $11{\times}11$ pixels is used.

\section{Conclusions and outlook}
We have developed an automated, flexible, modular, and robust TESS pipeline, optimized for producing light curves for asteroseismology from \acp{FFI} and \acp{TPF}, but useful for a wide variety of astrophysical applications. The pipeline runs in parallel at the Centre for Scientific Computing, Aarhus and is easily capable of keeping pace with monthly TESS data deliveries. Initial inputs are TESS FFIs and TPFs after low-level calibration has been performed by the SPOC. We use the TIC to identify all stars down to approximately Tmag = 15, estimate and remove time-resolved sky background from the images in a way which captures the complexity of the observed TESS backgrounds, and calculate precise timing information, including barycentric corrections, for all our targets.

After these processing steps, we then perform photometric extractions for all targets. The pipeline is designed for flexibility, and is able to transparently make use of multiple extraction algorithms including aperture photometry, two different PSF fitting implementations, difference imaging, and halo photometry. At present the default for most stars is aperture photometry, using an approach modeled on the K2P$^2$ algorithm developed for K2 data \citep{K2P2}. This methodology uses a clustering approach to assign pixels to groups of nearby targets and then segments the clusters to construct non-circular apertures for each individual star, allowing successful application of aperture photometry to even quite crowded fields. For the brightest stars (typically with $\Tmag \lesssim 4$) we apply an implementation of halo photometry \citep{halo,halopope} to construct a calibrated light curve from unsaturated pixels in the outer portions of the PSF of a saturated stellar image. The modular nature of our approach allows us to easily implement additional photometric algorithms, and we plan to include PSF fitting and difference imaging for crowded fields at a later stage. We have designed data formats for maximum compatibility with existing TESS products hosted at MAST and software widely used by the community such as \code{Lightkurve} \citep{lightkurve}. Noise metrics for the resulting light curves, whether measured using RMS or MDV, are generally consistent with expectations from the relations shown in \citet{Sullivan2015}. 

As noted above, we have designed the TASOC photometric pipeline with flexibility in mind. During the extended mission, the long cadence will be reduced to \LCC, while an additional short cadence mode of \SCC will be available on a limited shared-risk basis. Our pipeline is already able to successfully process these higher-cadence frames with limited impact on processing time, though we are still understanding the impacts of cosmic rays and potentially more limited calibration information available at these cadences. We anticipate continuing to make improvements to the pipeline and its products throughout the lifetime of the mission and beyond in response both to shortcomings we identify and to community input.

\begin{acknowledgements}
Funding for the Stellar Astrophysics Centre is provided by The Danish National Research Foundation (Grant agreement no.: DNRF106). 
RH and MNL acknowledges the ESA PRODEX programme.
This research was supported by the National Aeronautics and Space Administration (80NSSC18K1585 and 80NSSC19K0379) awarded through the TESS Guest Investigator Program.
OJH acknowledges the support of the UK Science and Technology Facilities Council (STFC).
This work was performed in part under contract with the Jet Propulsion Laboratory (JPL) funded by NASA through the Sagan Fellowship Program executed by the NASA Exoplanet Science Institute.

The numerical results presented in this work were obtained at the Centre for Scientific Computing, Aarhus\footnote{\url{https://phys.au.dk/forskning/cscaa/}}.

Resources supporting this work were provided by the NASA High-End Computing (HEC) Program through the NASA Advanced Supercomputing (NAS) Division at Ames Research Center for the production of the SPOC data products.

This research made use of Astropy,\footnote{\url{https://www.astropy.org}} a community-developed core Python package for Astronomy \citep{astropy:2013,astropy:2018}.

This paper includes data collected by the TESS mission, which are publicly available from the Mikulski Archive for Space Telescopes (MAST) and described in \citet{Jenkins2016}. Funding for the TESS mission is provided by NASA's Science Mission directorate. This research has made use of NASA's Astrophysics Data System, as well as the NASA/IPAC Extragalactic Database (NED) which is operated by the Jet Propulsion Laboratory, California Institute of Technology, under contract with the National Aeronautics and Space Administration.

Funding for the TESS Asteroseismic Science Operations Centre is provided by the Danish National Research Foundation (Grant agreement no.: DNRF106), ESA PRODEX (PEA 4000119301) and Stellar Astrophysics Centre (SAC) at Aarhus University. We thank the TESS team and staff and TASC/TASOC for their support of the present work.
\end{acknowledgements}

\software{Astropy \citep{astropy:2013, astropy:2018}, Photutils \citep{photutils}, NumPy \citep{numpy2020}, SciPy \citep{scipy}, matplotlib \citep{matplotlib}, scikit-learn \citep{scikit-learn}, scikit-image \citep{scikit-image}, \href{https://bottleneck.readthedocs.io}{Bottleneck}, SpiceyPy \citep{spiceypy}, \href{https://www.h5py.org}{h5py}, Q3C \citep{Koposov2006}.}

\facilities{TESS}


\appendix
\section{Background Tests}\label{app:background_tests}
TESS \acp{FFI} are contaminated by light from background sources (\eg Zodiacal light, scattered light from Solar System objects). This is of particular concern for TESS as it lies on an orbit that takes it close to earth periodically, causing large-scale changes in background flux.

Before settling on the methodology presented in \sref{subsec:background_estimation}, a series of background estimation algorithms were developed and tested on TESS data, both real and simulated \citep[\eg by][]{Lund2017}, and assessed on their ability to remove the background. The main difficulty for these methods is dealing with the corner glow, due to scattered light from stars off the CCD. Because the corner glow does not flow continuously into the rest of the background, methods relying on smoothing or interpolation tend to struggle slightly more.

\subsection{Row-by-column estimate}
The background continuum was estimated individually along each row of pixels in the FFI, and consequently along each column. Each column or row of data was divided into 60 bins. The 10th percentile value in the bins were then smoothed using a Hamming filter to provide a background continuum for a single row or column. The full FFI background was then obtained by averaging the two. This estimate performs well on the center of the CCD but under-corrects at the edges.

\subsection{Binned Mode estimate}
First, the data were subdivided into tiles of $128\!\times\!128$ pixels. The continuum pixel fluxes in each tile were fit using a RANSAC algorithm \citep{scikit-learn}, which provides outlier flags under the assumption that there are more background pixels than source pixels in each tile. The modes of the RANSAC-qualified values is taken as representative of that tile, resulting in 256 modes across the full FFI. These are then again fit with RANSAC. Finally, to obtain the background continuum, a 2D polynomial function is fit to the 256 modes using the RANSAC inlier probabilities as weights, thus minimizing the impact of tiles with increased density of stars resulting in anomalously high mode values. This method performs most poorly, as the RANSAC fitting is computationally expensive, and the spatial resolution does not effectively treat corner glow.

\subsection{\Kepler-style estimate}
The \Kepler mission \citep{Borucki2010} estimated the sky background across its camera by measuring flux in small $2\!\times\!2$ pixel tiles across each readout channel, in order to minimize contamination from recognizable stars, and then interpolating between them \citep{Bryson2010,KDPH-Bryson}.

This method attempts to do the same by taking tiles of $9\!\times\!9$ pixels equally distributed across the CCD and calculating the mode of the pixels inside these tiles, before interpolating between these data to obtain the background estimate for the full FFI. Interpolation tends to fail near the edges of the measured data, so in order to provide an accurate measure of the background the density of the points is doubled near the edges and corners of the FFI. The points are taken as the intersections of an irregular Cartesian grid, as was the original method for the \Kepler background. The modes were sigma clipped once at $1\sigma$, and then interpolated across the full FFI to obtain the background estimate. This method fails to consistently return a background measure, as the location of the pixels used to measure the background flux do not account for the location of any known stars, causing some stellar flux to be included in the background, and the interpolation can fail in frames with particularly bad corner glow (see \fref{fig:backgroundtests}). It should be stressed that this method is not the same as was used by the \Kepler mission, but merely inspired by it. In the \Kepler mission the FFIs from Kepler and the catalog were analyzed in order to define the locations of the 2x2 background apertures so that they accurately reflected the dimmest, normal pixels in each scene. In the version used here, this was not done.

\subsection{SE-style estimate}
The full $2048\!\times\!2048$ pixel image was cut into $64\!\times\!64$ pixel tiles, wherein a $3\sigma$ sigma clipping is performed to further mask out stars. The mode of the image was estimated using the same prescription as is used by SE \citep{SExtractor} (see \eqref{eq:background}, \sref{subsec:background_estimation}). It treats the corner glow by reparameterizing the image into radial coordinates, and performing the background fit a second time, but binned in radius instead of in cartesian coordinates. The two measurements are performed iteratively. These values were run through a $3\!\times\!3$ moving median filter and interpolated back onto the full image size.

\subsection{Method Comparison}
All four methods were used to estimate a simulated background. The `binned mode' estimate performed well in general but struggled in cases of significant corner glow, and had a large computational time due to the RANSAC fitting process.

The `row-by-column' estimate performed quickly and estimated the background well overall, but struggled near the edges due to smoothing methods. In some cases it introduced a regular grid-like structure due to the nature of the estimation method.

The `Kepler-style' method, failed to consistently produce smooth interpolations. Increasing the grid density increased the computational time to impractical length, and replacing the interpolation with a 2D polynomial fit failed to capture the global structure as well.

The `SE-style' estimate also performed quickly and estimated the background well overall, and is the only method that sufficiently treats the corner glow. Due to the robustness and simplicity of the underlying code and methodology, this method was applied in the main pipeline.

\begin{figure*}[h!]
    \centering
    \includegraphics[width=.95\textwidth]{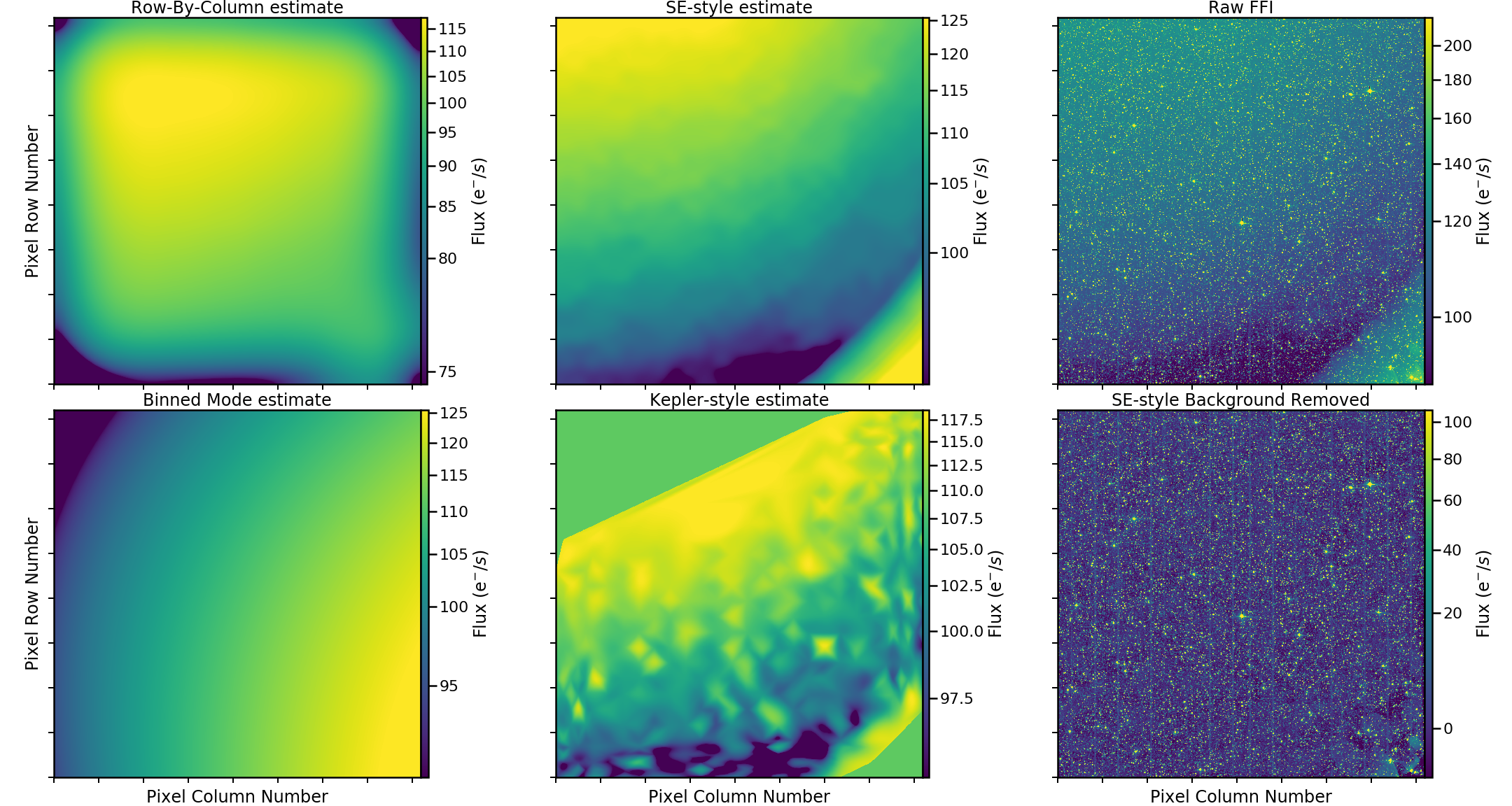}
    \caption{Background estimates of a FFI from TESS Camera 1, CCD 2 in Sector 3, for each of the four estimation methods. The raw and background-subtracted image are shown on the right. The SE-style estimate removes the background most robustly.}
    \label{fig:backgroundtests}
\end{figure*}

The four background estimates, along with raw and SE-style background subtracted FFIs, are shown in \fref{fig:backgroundtests}.

\FloatBarrier
\onecolumngrid
\section{FITS headers of output light curves}
We here include an example of the full FITS headers of a typical FITS light curve file produced by the TASOC Photometry Pipeline as described in \sref{sec:data_formats}. In extensions 3 and 4 (\code{SUMIMAGE} and \code{APERTURE}), several headers pertaining to the World Coordinate System solution and SIP distortions have been removed to save space.
\subsection*{HDU-0 (PRIMARY)}
\begin{Verbatim}[fontsize=\small]
SIMPLE  =                    T / conforms to FITS standard
BITPIX  =                    8 / array data type
NAXIS   =                    0 / number of array dimensions
EXTEND  =                    T
NEXTEND =                    3 / number of standard extensions
EXTNAME = 'PRIMARY '           / name of extension
ORIGIN  = 'TASOC/Aarhus'       / institution responsible for creating this file
DATE    = '2020-06-06'         / date the file was created
TELESCOP= 'TESS    '           / telescope
INSTRUME= 'TESS Photometer'    / detector type
FILTER  = 'TESS    '           / Photometric bandpass filter
OBJECT  = 'TIC 444953480'      / string version of TICID
TICID   =            444953480 / unique TESS target identifier
CAMERA  =                    2 / Camera number
CCD     =                    4 / CCD number
SECTOR  =                    6 / Observing sector
PROCVER = 'master-v4.5'        / Version of photometry pipeline
FILEVER = '1.4     '           / File format version
DATA_REL=                    8 / Data release number
VERSION =                    5 / Version of the processing
PHOTMET = 'aperture'           / Photometric method used
RADESYS = 'ICRS    '           / reference frame of celestial coordinates
EQUINOX =               2000.0 / equinox of celestial coordinate system
RA_OBJ  =            91.263592 / [deg] Right ascension
DEC_OBJ =           -12.035933 / [deg] Declination
PMRA    =   0.9823499999999999 / [mas/yr] RA proper motion
PMDEC   =            -0.199674 / [mas/yr] Dec proper motion
PMTOTAL =    1.002437643335484 / [mas/yr] total proper motion
TESSMAG =               13.668 / [mag] TESS magnitude
TEFF    =               4878.0 / [K] Effective temperature
TICVER  =                    7 / TESS Input Catalog version
CRMITEN =                    T / spacecraft cosmic ray mitigation enabled
CRBLKSZ =                   10 / [exposures] s/c cosmic ray mitigation block siz
CRSPOC  =                    F / SPOC cosmic ray cleaning enabled
KP_THRES=                  0.8 / K2P2 sum-image threshold
KP_MIPIX=                    4 / K2P2 min pixels in mask
KP_MICLS=                    4 / K2P2 min pix. for cluster
KP_CLSRA=    1.414213562373095 / K2P2 cluster radius
KP_WS   =                    T / K2P2 watershed segmentation
KP_WSBLR=                  0.5 / K2P2 watershed blur
KP_WSTHR=                    0 / K2P2 watershed threshold
KP_WSFOT=                    3 / K2P2 watershed footprint
KP_EX   =                    T / K2P2 extend overflow
AP_CONT =   0.5487687205693473 / AP contamination
DATAVAL =                    0 / Data validation flags
CHECKSUM= 'geOejdNegdNegdNe'   / HDU checksum updated 2020-06-06T16:44:04
DATASUM = '0       '           / data unit checksum updated 2020-06-06T16:44:04
\end{Verbatim}

\subsection*{HDU-1 (LIGHTCURVE)}
\begin{Verbatim}[fontsize=\small]
XTENSION= 'BINTABLE'           / binary table extension
BITPIX  =                    8 / array data type
NAXIS   =                    2 / number of array dimensions
NAXIS1  =                   96 / length of dimension 1
NAXIS2  =                  993 / length of dimension 2
PCOUNT  =                    0 / number of group parameters
GCOUNT  =                    1 / number of groups
TFIELDS =                   14 / number of table fields
INHERIT =                    T / inherit the primary header
TTYPE1  = 'TIME    '           / column title: data time stamps
TFORM1  = 'D       '           / column format: 64-bit floating point
TUNIT1  = 'BJD - 2457000, days' / column units: Barycenter corrected TESS Julian
TDISP1  = 'D14.7   '           / column display format
TTYPE2  = 'TIMECORR'           / column title: barycenter - timeslice correction
TFORM2  = 'E       '           / column format: 32-bit floating point
TUNIT2  = 'd       '           / column units: day
TDISP2  = 'E13.6   '           / column display format
TTYPE3  = 'CADENCENO'          / column title: unique cadence number
TFORM3  = 'J       '           / column format: signed 32-bit integer
TDISP3  = 'I10     '           / column display format
TTYPE4  = 'FLUX_RAW'           / column title: photometric flux
TFORM4  = 'D       '           / column format: 64-bit floating point
TUNIT4  = 'e-/s    '           / column units: electrons per second
TDISP4  = 'E26.17  '           / column display format
TTYPE5  = 'FLUX_RAW_ERR'       / column title: photometric flux error
TFORM5  = 'D       '           / column format: 64-bit floating point
TUNIT5  = 'e-/s    '           / column units: electrons per second
TDISP5  = 'E26.17  '           / column display format
TTYPE6  = 'FLUX_BKG'           / column title: photometric background flux
TFORM6  = 'D       '           / column format: 64-bit floating point
TUNIT6  = 'e-/s    '           / column units: electrons per second
TDISP6  = 'E26.17  '           / column display format
TTYPE7  = 'FLUX_CORR'          / column title: corrected photometric flux
TFORM7  = 'D       '           / column format: 64-bit floating point
TUNIT7  = 'ppm     '           / column units: rel. flux in parts-per-million
TDISP7  = 'E26.17  '           / column display format
TTYPE8  = 'FLUX_CORR_ERR'      / column title: corrected photometric flux error
TFORM8  = 'D       '           / column format: 64-bit floating point
TUNIT8  = 'ppm     '           / column units: parts-per-million
TDISP8  = 'E26.17  '           / column display format
TTYPE9  = 'QUALITY '           / column title: photometry quality flags
TFORM9  = 'J       '           / column format: signed 32-bit integer
TDISP9  = 'B16.16  '           / column display format
TTYPE10 = 'PIXEL_QUALITY'      / column title: pixel quality flags
TFORM10 = 'J       '           / column format: signed 32-bit integer
TDISP10 = 'B16.16  '           / column display format
TTYPE11 = 'MOM_CENTR1'         / column title: moment-derived column centroid
TFORM11 = 'D       '           / column format: 64-bit floating point
TUNIT11 = 'pixels  '           / column units: pixels
TDISP11 = 'F10.5   '           / column display format
TTYPE12 = 'MOM_CENTR2'         / column title: moment-derived row centroid
TFORM12 = 'D       '           / column format: 64-bit floating point
TUNIT12 = 'pixels  '           / column units: pixels
TDISP12 = 'F10.5   '           / column display format
TTYPE13 = 'POS_CORR1'          / column title: column position correction
TFORM13 = 'D       '           / column format: 64-bit floating point
TUNIT13 = 'pixels  '           / column units: pixels
TDISP13 = 'F14.7   '           / column display format
TTYPE14 = 'POS_CORR2'          / column title: row position correction
TFORM14 = 'D       '           / column format: 64-bit floating point
TUNIT14 = 'pixels  '           / column units: pixels
TDISP14 = 'F14.7   '           / column display format
EXTNAME = 'LIGHTCURVE'         / extension name
TIMEREF = 'SOLARSYSTEM'        / barycentric correction applied to times
TIMESYS = 'TDB     '           / time system is Barycentric Dynamical Time (TDB)
BJDREFI =              2457000 / integer part of BTJD reference date
BJDREFF =                  0.0 / fraction of the day in BTJD reference date
TIMEUNIT= 'd       '           / time unit for TIME, TSTART and TSTOP
TSTART  =    1468.276025411983 / observation start time in BTJD
TSTOP   =    1490.046743795897 / observation stop time in BTJD
DATE-OBS= '2018-12-15T18:36:19.412' / TSTART as UTC calendar date
DATE-END= '2019-01-06T13:06:09.480' / TSTOP as UTC calendar date
MJD-BEG =    58467.77602541199 / observation start time in MJD
MJD-END =    58489.54674379589 / observation start time in MJD
TELAPSE =    21.77071838391339 / [d] TSTOP - TSTART
LIVETIME=    4.310602240014852 / [d] TELAPSE multiplied by DEADC
DEADC   =                0.198 / deadtime correction
EXPOSURE=    4.310602240014852 / [d] time on source
XPOSURE =                356.4 / [s] Duration of exposure
TIMEPIXR=                  0.5 / bin time beginning=0 middle=0.5 end=1
TIMEDEL =  0.02083333333333333 / [d] time resolution of data
INT_TIME=                 1.98 / [s] photon accumulation time per frame
READTIME=                 0.02 / [s] readout time per frame
FRAMETIM=                  2.0 / [s] frame time (INT_TIME + READTIME)
NUM_FRM =                  900 / number of frames per time stamp
NREADOUT=                  720 / number of read per cadence
CHECKSUM= 'eeSbfZSbedSbeZSb'   / HDU checksum updated 2020-06-06T16:44:04
DATASUM = '3537856224'         / data unit checksum updated 2020-06-06T16:44:04
\end{Verbatim}

\subsection*{HDU-2 (SUMIMAGE)}
\begin{Verbatim}[fontsize=\small]
XTENSION= 'IMAGE   '           / Image extension
BITPIX  =                  -64 / array data type
NAXIS   =                    2 / number of array dimensions
NAXIS1  =                   15
NAXIS2  =                   15
PCOUNT  =                    0 / number of parameters
GCOUNT  =                    1 / number of groups
INHERIT =                    T / inherit the primary header
WCSAXES =                    2 / Number of coordinate axes
CRPIX1  =                863.0 / Pixel coordinate of reference point
CRPIX2  =                 18.0 / Pixel coordinate of reference point
PC1_1   =  -0.0057098640987826 / Coordinate transformation matrix element
PC1_2   =  -0.0003675563609303 / Coordinate transformation matrix element
PC2_1   =  0.00022773533217156 / Coordinate transformation matrix element
PC2_2   =   -0.005696929639529 / Coordinate transformation matrix element
CDELT1  =                  1.0 / [deg] Coordinate increment at reference point
CDELT2  =                  1.0 / [deg] Coordinate increment at reference point
CUNIT1  = 'deg'                / Units of coordinate increment and value
CUNIT2  = 'deg'                / Units of coordinate increment and value
CTYPE1  = 'RA---TAN-SIP'       / TAN (gnomonic) projection + SIP distortions
CTYPE2  = 'DEC--TAN-SIP'       / TAN (gnomonic) projection + SIP distortions
CRVAL1  =       86.20939522998 / [deg] Coordinate value at reference point
CRVAL2  =     -11.930165289899 / [deg] Coordinate value at reference point
LONPOLE =                180.0 / [deg] Native longitude of celestial pole
LATPOLE =     -11.930165289899 / [deg] Native latitude of celestial pole
TIMESYS = 'TDB'                / Time scale
TIMEUNIT= 'd'                  / Time units
DATEREF = '1858-11-17'         / ISO-8601 fiducial time
MJDREFI =                  0.0 / [d] MJD of fiducial time, integer part
MJDREFF =                  0.0 / [d] MJD of fiducial time, fractional part
DATE-OBS= '2018-12-19T00:06:23.366Z' / ISO-8601 time of observation
MJD-OBS =      58471.004437106 / [d] MJD of observation
MJD-OBS =      58471.004437106 / [d] MJD at start of observation
TSTART  =      1471.5052378466 / [d] Time elapsed since fiducial time at start
DATE-END= '2018-12-19T00:36:23.366Z' / ISO-8601 time at end of observation
MJD-END =       58471.02527044 / [d] MJD at end of observation
TSTOP   =      1471.5260711749 / [d] Time elapsed since fiducial time at end
TELAPSE =       0.020833328235 / [d] Elapsed time (start to stop)
TIMEDEL =    0.020833333333333 / [d] Time resolution
TIMEPIXR=                  0.5 / Reference position of timestamp in binned data
RADESYS = 'ICRS'               / Equatorial coordinate system
EXTNAME = 'SUMIMAGE'           / extension name
CHECKSUM= 'HS9hIP8fHP8fHP8f'   / HDU checksum updated 2020-06-06T16:44:04
DATASUM = '2068211304'         / data unit checksum updated 2020-06-06T16:44:04
\end{Verbatim}

\subsection*{HDU-3 (APERTURE)}
\begin{Verbatim}[fontsize=\small]
XTENSION= 'IMAGE   '           / Image extension
BITPIX  =                   32 / array data type
NAXIS   =                    2 / number of array dimensions
NAXIS1  =                   15
NAXIS2  =                   15
PCOUNT  =                    0 / number of parameters
GCOUNT  =                    1 / number of groups
INHERIT =                    T / inherit the primary header
WCSAXES =                    2 / Number of coordinate axes
CRPIX1  =                863.0 / Pixel coordinate of reference point
CRPIX2  =                 18.0 / Pixel coordinate of reference point
PC1_1   =  -0.0057098640987826 / Coordinate transformation matrix element
PC1_2   =  -0.0003675563609303 / Coordinate transformation matrix element
PC2_1   =  0.00022773533217156 / Coordinate transformation matrix element
PC2_2   =   -0.005696929639529 / Coordinate transformation matrix element
CDELT1  =                  1.0 / [deg] Coordinate increment at reference point
CDELT2  =                  1.0 / [deg] Coordinate increment at reference point
CUNIT1  = 'deg'                / Units of coordinate increment and value
CUNIT2  = 'deg'                / Units of coordinate increment and value
CTYPE1  = 'RA---TAN-SIP'       / TAN (gnomonic) projection + SIP distortions
CTYPE2  = 'DEC--TAN-SIP'       / TAN (gnomonic) projection + SIP distortions
CRVAL1  =       86.20939522998 / [deg] Coordinate value at reference point
CRVAL2  =     -11.930165289899 / [deg] Coordinate value at reference point
LONPOLE =                180.0 / [deg] Native longitude of celestial pole
LATPOLE =     -11.930165289899 / [deg] Native latitude of celestial pole
TIMESYS = 'TDB'                / Time scale
TIMEUNIT= 'd'                  / Time units
DATEREF = '1858-11-17'         / ISO-8601 fiducial time
MJDREFI =                  0.0 / [d] MJD of fiducial time, integer part
MJDREFF =                  0.0 / [d] MJD of fiducial time, fractional part
DATE-OBS= '2018-12-19T00:06:23.366Z' / ISO-8601 time of observation
MJD-OBS =      58471.004437106 / [d] MJD of observation
MJD-OBS =      58471.004437106 / [d] MJD at start of observation
TSTART  =      1471.5052378466 / [d] Time elapsed since fiducial time at start
DATE-END= '2018-12-19T00:36:23.366Z' / ISO-8601 time at end of observation
MJD-END =       58471.02527044 / [d] MJD at end of observation
TSTOP   =      1471.5260711749 / [d] Time elapsed since fiducial time at end
TELAPSE =       0.020833328235 / [d] Elapsed time (start to stop)
TIMEDEL =    0.020833333333333 / [d] Time resolution
TIMEPIXR=                  0.5 / Reference position of timestamp in binned data
RADESYS = 'ICRS'               / Equatorial coordinate system
EXTNAME = 'APERTURE'           / extension name
CHECKSUM= 'U6QAW4P5U4PAU4P5'   / HDU checksum updated 2020-06-06T16:44:04
DATASUM = '8545    '           / data unit checksum updated 2020-06-06T16:44:04
\end{Verbatim}

\bibliographystyle{aasjournal}
\bibliography{biblio}

\begin{thebibliography}{}
\expandafter\ifx\csname natexlab\endcsname\relax\def\natexlab#1{#1}\fi
\providecommand{\url}[1]{\href{#1}{#1}}
\providecommand{\dodoi}[1]{doi:~\href{http://doi.org/#1}{\nolinkurl{#1}}}
\providecommand{\doeprint}[1]{\href{http://ascl.net/#1}{\nolinkurl{http://ascl.net/#1}}}
\providecommand{\doarXiv}[1]{\href{https://arxiv.org/abs/#1}{\nolinkurl{https://arxiv.org/abs/#1}}}

\bibitem[{{Aerts} {et~al.}(2010){Aerts}, {Christensen-Dalsgaard}, \&
  {Kurtz}}]{TheBook}
{Aerts}, C., {Christensen-Dalsgaard}, J., \& {Kurtz}, D.~W. 2010,
  {Asteroseismology} (Springer)

\bibitem[{Annex {et~al.}(2019)Annex, Carcich, kd7uiy, Badger, ya~Murakami,
  Kulumani, de~Val-Borro, Stefko, del Rio, \& Seignovert}]{spiceypy}
Annex, A., Carcich, B., kd7uiy, {et~al.} 2019, AndrewAnnex/SpiceyPy: SpiceyPy
  2.2.0, \dodoi{10.5281/zenodo.2576445}

\bibitem[{{Astropy Collaboration} {et~al.}(2013){Astropy Collaboration},
  {Robitaille}, {Tollerud}, {Greenfield}, {Droettboom}, {Bray}, {Aldcroft},
  {Davis}, {Ginsburg}, {Price-Whelan}, {Kerzendorf}, {Conley}, {Crighton},
  {Barbary}, {Muna}, {Ferguson}, {Grollier}, {Parikh}, {Nair}, {Unther},
  {Deil}, {Woillez}, {Conseil}, {Kramer}, {Turner}, {Singer}, {Fox}, {Weaver},
  {Zabalza}, {Edwards}, {Azalee Bostroem}, {Burke}, {Casey}, {Crawford},
  {Dencheva}, {Ely}, {Jenness}, {Labrie}, {Lim}, {Pierfederici}, {Pontzen},
  {Ptak}, {Refsdal}, {Servillat}, \& {Streicher}}]{astropy:2013}
{Astropy Collaboration}, {Robitaille}, T.~P., {Tollerud}, E.~J., {et~al.} 2013,
  \aap, 558, A33, \dodoi{10.1051/0004-6361/201322068}

\bibitem[{{Baglin} {et~al.}(2002){Baglin}, {Auvergne}, {Barge}, {Buey},
  {Catala}, {Michel}, {Weiss}, \& {COROT Team}}]{Baglin2002}
{Baglin}, A., {Auvergne}, M., {Barge}, P., {et~al.} 2002, in ESA Special
  Publication, Vol. 485, Stellar Structure and Habitable Planet Finding, ed.
  B.~{Battrick}, F.~{Favata}, I.~W. {Roxburgh}, \& D.~{Galadi}, 17--24

\bibitem[{{Batalha} {et~al.}(2011){Batalha}, {Borucki}, {Bryson}, {Buchhave},
  {Caldwell}, {Christensen-Dalsgaard}, {Ciardi}, {Dunham}, {Fressin},
  {Gautier}, {Gilliland}, {Haas}, {Howell}, {Jenkins}, {Kjeldsen}, {Koch},
  {Latham}, {Lissauer}, {Marcy}, {Rowe}, {Sasselov}, {Seager}, {Steffen},
  {Torres}, {Basri}, {Brown}, {Charbonneau}, {Christiansen}, {Clarke},
  {Cochran}, {Dupree}, {Fabrycky}, {Fischer}, {Ford}, {Fortney}, {Girouard},
  {Holman}, {Johnson}, {Isaacson}, {Klaus}, {Machalek}, {Moorehead},
  {Morehead}, {Ragozzine}, {Tenenbaum}, {Twicken}, {Quinn}, {VanCleve},
  {Walkowicz}, {Welsh}, {Devore}, \& {Gould}}]{Batalha2011}
{Batalha}, N.~M., {Borucki}, W.~J., {Bryson}, S.~T., {et~al.} 2011, \apj, 729,
  27, \dodoi{10.1088/0004-637X/729/1/27}

\bibitem[{{Berta-Thompson} {et~al.}(2015){Berta-Thompson}, {Levine}, \&
  {Sullivan}}]{CosmicRayMitigation}
{Berta-Thompson}, Z.~K., {Levine}, A., \& {Sullivan}, P. 2015, {Cosmic Ray
  Rejection Strategies for TESS}, Tech. rep., Kavli Institute for Astrophysics
  and Space Science, Massachusetts Institute of Technology

\bibitem[{Bertin \& Arnouts(1996)}]{SExtractor}
Bertin, E., \& Arnouts, S. 1996, Astron. Astrophys. Suppl. Ser., 117, 393,
  \dodoi{10.1051/aas:1996164}

\bibitem[{{Borucki} {et~al.}(2010){Borucki}, {Koch}, {Basri}, {Batalha},
  {Brown}, {Caldwell}, {Caldwell}, {Christensen-Dalsgaard}, {Cochran},
  {DeVore}, {Dunham}, {Dupree}, {Gautier}, {Geary}, {Gilliland}, {Gould},
  {Howell}, {Jenkins}, {Kondo}, {Latham}, {Marcy}, {Meibom}, {Kjeldsen},
  {Lissauer}, {Monet}, {Morrison}, {Sasselov}, {Tarter}, {Boss}, {Brownlee},
  {Owen}, {Buzasi}, {Charbonneau}, {Doyle}, {Fortney}, {Ford}, {Holman},
  {Seager}, {Steffen}, {Welsh}, {Rowe}, {Anderson}, {Buchhave}, {Ciardi},
  {Walkowicz}, {Sherry}, {Horch}, {Isaacson}, {Everett}, {Fischer}, {Torres},
  {Johnson}, {Endl}, {MacQueen}, {Bryson}, {Dotson}, {Haas}, {Kolodziejczak},
  {Van Cleve}, {Chandrasekaran}, {Twicken}, {Quintana}, {Clarke}, {Allen},
  {Li}, {Wu}, {Tenenbaum}, {Verner}, {Bruhweiler}, {Barnes}, \&
  {Prsa}}]{Borucki2010}
{Borucki}, W.~J., {Koch}, D., {Basri}, G., {et~al.} 2010, Science, 327, 977,
  \dodoi{10.1126/science.1185402}

\bibitem[{{Bouma} {et~al.}(2019){Bouma}, {Hartman}, {Bhatti}, {Winn}, \&
  {Bakos}}]{cdips}
{Bouma}, L.~G., {Hartman}, J.~D., {Bhatti}, W., {Winn}, J.~N., \& {Bakos},
  G.~{\'A}. 2019, \apjs, 245, 13, \dodoi{10.3847/1538-4365/ab4a7e}

\bibitem[{Bradley {et~al.}(2019)Bradley, Sipocz, Robitaille, Tollerud,
  Vinícius, Deil, Barbary, Günther, Cara, Busko, Conseil, Droettboom,
  Bostroem, Bray, Bratholm, Wilson, Craig, Barentsen, Pascual, Donath, Greco,
  Perren, Lim, \& Kerzendorf}]{photutils}
Bradley, L., Sipocz, B., Robitaille, T., {et~al.} 2019, astropy/photutils:
  v0.6, \dodoi{10.5281/zenodo.2533376}

\bibitem[{{Brasseur} {et~al.}(2019){Brasseur}, {Phillip}, {Fleming},
  {Mullally}, \& {White}}]{tesscut}
{Brasseur}, C.~E., {Phillip}, C., {Fleming}, S.~W., {Mullally}, S.~E., \&
  {White}, R.~L. 2019, {Astrocut: Tools for creating cutouts of TESS images}.
\newblock \doeprint{1905.007}

\bibitem[{{Bryson} {et~al.}(2010){Bryson}, {Jenkins}, {Klaus}, {Cote},
  {Quintana}, {Hall}, {Ibrahim}, {Chandrasekaran}, {Caldwell}, {Van Cleve}, \&
  {Haas}}]{Bryson2010}
{Bryson}, S.~T., {Jenkins}, J.~M., {Klaus}, T.~C., {et~al.} 2010, in Society of
  Photo-Optical Instrumentation Engineers (SPIE) Conference Series, Vol. 7740,
  Software and Cyberinfrastructure for Astronomy, 77401D,
  \dodoi{10.1117/12.857625}

\bibitem[{{Bryson} {et~al.}(2020){Bryson}, {Jenkins}, {Klaus}, {Cote},
  {Quintana}, {Campbell}, {Zamudio}, {Chandrasekaran}, {Caldwell}, {Van Cleve},
  \& {Haas}}]{KDPH-Bryson}
{Bryson}, S.~T., {Jenkins}, J.~M., {Klaus}, T.~C., {et~al.} 2020, {Kepler Data
  Processing Handbook: Target and Aperture Definitions: Selecting Pixels for
  Kepler Downlink}, Kepler Data Processing Handbook (KSCI-19081-003)

\bibitem[{{Caldwell} {et~al.}(2020){Caldwell}, {Tenenbaum}, {Twicken},
  {Jenkins}, {Ting}, {Smith}, {Hedges}, {Fausnaugh}, {Rose}, \&
  {Burke}}]{Caldwell2020}
{Caldwell}, D.~A., {Tenenbaum}, P., {Twicken}, J.~D., {et~al.} 2020, Research
  Notes of the American Astronomical Society, 4, 201,
  \dodoi{10.3847/2515-5172/abc9b3}

\bibitem[{{Clarke} {et~al.}(2020){Clarke}, {Caldwell}, {Quintana},
  {Chandrasekaran}, {Twicken}, {Jenkins}, {Cote}, {McCauliff}, {Klaus},
  {Allen}, \& {Bryson}}]{KDPH-pixel}
{Clarke}, B.~D., {Caldwell}, D.~A., {Quintana}, E.~V., {et~al.} 2020, {Kepler
  Data Processing Handbook: Pixel Level Calibrations}, Kepler Science Document
  KSCI-19081-003

\bibitem[{{Eisner} {et~al.}(2019){Eisner}, {Pope}, {Aigrain}, {Barrag{\'a}n},
  {White}, {Huang}, {Lintott}, \& {Volkov}}]{eisner19}
{Eisner}, N.~L., {Pope}, B. J.~S., {Aigrain}, S., {et~al.} 2019, Research Notes
  of the American Astronomical Society, 3, 145,
  \dodoi{10.3847/2515-5172/ab49ff}

\bibitem[{Ester {et~al.}(1996)Ester, Kriegel, Sander, \& Xu}]{dbscan}
Ester, M., Kriegel, H.-P., Sander, J., \& Xu, X. 1996, in Proceedings of the
  Second International Conference on Knowledge Discovery and Data Mining,
  KDD'96 (AAAI Press), 226--231, \dodoi{10.5555/3001460.3001507}

\bibitem[{{Fausnaugh} {et~al.}(2018){Fausnaugh}, {Huang}, {Glidden},
  {Guerrero}, \& {TESS Science Office}}]{quicklookpipeline}
{Fausnaugh}, M., {Huang}, X., {Glidden}, A., {Guerrero}, N., \& {TESS Science
  Office}. 2018, in American Astronomical Society Meeting Abstracts, Vol. 231,
  American Astronomical Society Meeting Abstracts \#231, 439.09

\bibitem[{Fausnaugh {et~al.}(2019)Fausnaugh, Burke, Caldwell, Jenkins, Smith,
  Twicken, Vanderspek, Doty, Ting, \& Villasenor}]{tessdr25}
Fausnaugh, M.~M., Burke, C.~J., Caldwell, D.~A., {et~al.} 2019, TESS Data
  Release Notes: Sector 18, DR25, NASA.
\newblock
  \url{https://archive.stsci.edu/missions/tess/doc/tess_drn/tess_sector_18_drn25_v02.pdf}

\bibitem[{Fausnaugh {et~al.}(2020{\natexlab{a}})Fausnaugh, Burke, Caldwell,
  Jenkins, Smith, Twicken, Vanderspek, Doty, Ting, \&
  Villasenor}]{tessdr27memo}
---. 2020{\natexlab{a}}, TESS Data Release Notes 27 Memo: Updates to Sector 20,
  Data Release 27 Products, NASA.
\newblock
  \url{https://archive.stsci.edu/missions/tess/doc/tess_drn/tess_s20_dr27_data_product_revision_memo_v02.pdf}

\bibitem[{Fausnaugh {et~al.}(2020{\natexlab{b}})Fausnaugh, Burke, Caldwell,
  Jenkins, Smith, Twicken, Vanderspek, Doty, Ting, \&
  Villasenor}]{tessdr29memo}
---. 2020{\natexlab{b}}, TESS Data Release Notes 29 Memo: Updates to Sector 21,
  Data Release 29 Products, NASA.
\newblock
  \url{https://archive.stsci.edu/missions/tess/doc/tess_drn/tess_s21_dr29_data_product_revision_memo_v02.pdf}

\bibitem[{{Feinstein} {et~al.}(2019){Feinstein}, {Montet}, {Foreman-Mackey},
  {Bedell}, {Saunders}, {Bean}, {Christiansen}, {Hedges}, {Luger}, {Scolnic},
  \& {Cardoso}}]{eleanor}
{Feinstein}, A.~D., {Montet}, B.~T., {Foreman-Mackey}, D., {et~al.} 2019,
  \pasp, 131, 094502, \dodoi{10.1088/1538-3873/ab291c}

\bibitem[{{Gaia Collaboration} {et~al.}(2018){Gaia Collaboration}, Brown,
  Vallenari, Prusti, de~Bruijne, Babusiaux, \& Bailer-Jones}]{GaiaDR2}
{Gaia Collaboration}, Brown, A. G.~A., Vallenari, A., {et~al.} 2018, Astron.
  Astrophys., 616, A1, \dodoi{10.1051/0004-6361/201833051}

\bibitem[{{Handberg} \& {Lund}(2014)}]{kasoc_filt}
{Handberg}, R., \& {Lund}, M.~N. 2014, \mnras, 445, 2698,
  \dodoi{10.1093/mnras/stu1823}

\bibitem[{Handberg \& Lund(2017)}]{Handberg2017}
Handberg, R., \& Lund, M.~N. 2017, Astron. Astrophys., 597, A36,
  \dodoi{10.1051/0004-6361/201527753}

\bibitem[{Harris {et~al.}(2020)Harris, Millman, van~der Walt, Gommers,
  Virtanen, Cournapeau, Wieser, Taylor, Berg, Smith, Kern, Picus, Hoyer, van
  Kerkwijk, Brett, Haldane, del R{'{\i}}o, Wiebe, Peterson,
  G{'{e}}rard-Marchant, Sheppard, Reddy, Weckesser, Abbasi, Gohlke, \&
  Oliphant}]{numpy2020}
Harris, C.~R., Millman, K.~J., van~der Walt, S.~J., {et~al.} 2020, Nature, 585,
  357, \dodoi{10.1038/s41586-020-2649-2}

\bibitem[{{Howell} {et~al.}(2014){Howell}, {Sobeck}, {Haas}, {Still},
  {Barclay}, {Mullally}, {Troeltzsch}, {Aigrain}, {Bryson}, {Caldwell},
  {Chaplin}, {Cochran}, {Huber}, {Marcy}, {Miglio}, {Najita}, {Smith},
  {Twicken}, \& {Fortney}}]{Howell2014}
{Howell}, S.~B., {Sobeck}, C., {Haas}, M., {et~al.} 2014, \pasp, 126, 398,
  \dodoi{10.1086/676406}

\bibitem[{{Huang} {et~al.}(2020){Huang}, {Vanderburg}, {P{\'a}l}, {Sha}, {Yu},
  {Fong}, {Fausnaugh}, {Shporer}, {Guerrero}, {Vanderspek}, \&
  {Ricker}}]{Huang2020a}
{Huang}, C.~X., {Vanderburg}, A., {P{\'a}l}, A., {et~al.} 2020, Research Notes
  of the American Astronomical Society, 4, 204,
  \dodoi{10.3847/2515-5172/abca2e}

\bibitem[{Hunter(2007)}]{matplotlib}
Hunter, J.~D. 2007, Computing in Science \& Engineering, 9, 90,
  \dodoi{10.1109/MCSE.2007.55}

\bibitem[{{Jenkins}(2017)}]{Jenkins2017}
{Jenkins}, J.~M. 2017, {Kepler Data Processing Handbook: KSCI-19081-002}

\bibitem[{{Jenkins} {et~al.}(2016){Jenkins}, {Twicken}, {McCauliff},
  {Campbell}, {Sanderfer}, {Lung}, {Mansouri-Samani}, {Girouard}, {Tenenbaum},
  {Klaus}, {Smith}, {Caldwell}, {Chacon}, {Henze}, {Heiges}, {Latham},
  {Morgan}, {Swade}, {Rinehart}, \& {Vanderspek}}]{Jenkins2016}
{Jenkins}, J.~M., {Twicken}, J.~D., {McCauliff}, S., {et~al.} 2016, in Society
  of Photo-Optical Instrumentation Engineers (SPIE) Conference Series, Vol.
  9913, \procspie, 99133E, \dodoi{10.1117/12.2233418}

\bibitem[{Kjeldsen \& Frandsen(1992)}]{MOMF}
Kjeldsen, H., \& Frandsen, S. 1992, Publ. Astron. Soc. Pacific, 104, 413,
  \dodoi{10.1086/133014}

\bibitem[{Koposov \& Bartunov(2006)}]{Koposov2006}
Koposov, S., \& Bartunov, O. 2006, Astron. Data Anal. Softw. Syst. XV, 351,
  735.
\newblock \url{http://adsabs.harvard.edu/full/2006ASPC..351..735K}

\bibitem[{{Lightkurve Collaboration} {et~al.}(2018){Lightkurve Collaboration},
  {Cardoso}, {Hedges}, {Gully-Santiago}, {Saunders}, {Cody}, {Barclay}, {Hall},
  {Sagear}, {Turtelboom}, {Zhang}, {Tzanidakis}, {Mighell}, {Coughlin}, {Bell},
  {Berta-Thompson}, {Williams}, {Dotson}, \& {Barentsen}}]{lightkurve}
{Lightkurve Collaboration}, {Cardoso}, J.~V.~d.~M., {Hedges}, C., {et~al.}
  2018, {Lightkurve: Kepler and TESS time series analysis in Python},
  Astrophysics Source Code Library.
\newblock \doeprint{1812.013}

\bibitem[{{Luger} {et~al.}(2016){Luger}, {Agol}, {Kruse}, {Barnes}, {Becker},
  {Foreman-Mackey}, \& {Deming}}]{Luger2016}
{Luger}, R., {Agol}, E., {Kruse}, E., {et~al.} 2016, \aj, 152, 100,
  \dodoi{10.3847/0004-6256/152/4/100}

\bibitem[{Lund {et~al.}(2015)Lund, Handberg, Davies, Chaplin, \& Jones}]{K2P2}
Lund, M.~N., Handberg, R., Davies, G.~R., Chaplin, W.~J., \& Jones, C.~D. 2015,
  Astrophys. J., 806, 30, \dodoi{10.1088/0004-637X/806/1/30}

\bibitem[{Lund {et~al.}(2017)Lund, Handberg, Kjeldsen, Chaplin, \&
  Christensen-Dalsgaard}]{Lund_Azores2017}
Lund, M.~N., Handberg, R., Kjeldsen, H., Chaplin, W.~J., \&
  Christensen-Dalsgaard, J. 2017, EPJ Web Conf., 160, 01005,
  \dodoi{10.1051/epjconf/201716001005}

\bibitem[{{Lund} {et~al.}(2017){Lund}, {Handberg}, {Kjeldsen}, {Chaplin}, \&
  {Christensen-Dalsgaard}}]{Lund2017}
{Lund}, M.~N., {Handberg}, R., {Kjeldsen}, H., {Chaplin}, W.~J., \&
  {Christensen-Dalsgaard}, J. 2017, EPJ Web Conf., 160, 01005,
  \dodoi{10.1051/epjconf/201716001005}

\bibitem[{Maclaurin {et~al.}(2015)Maclaurin, Duvenaud, \& Adams}]{autograd}
Maclaurin, D., Duvenaud, D., \& Adams, R.~P. 2015, in ICML 2015 AutoML Workshop

\bibitem[{{Miglio} {et~al.}(2013){Miglio}, {Chiappini}, {Morel}, {Barbieri},
  {Chaplin}, {Girardi}, {Montalb{\'a}n}, {Valentini}, {Mosser}, {Baudin},
  {Casagrande}, {Fossati}, {Silva Aguirre}, \& {Baglin}}]{Miglio2013}
{Miglio}, A., {Chiappini}, C., {Morel}, T., {et~al.} 2013, \mnras, 429, 423,
  \dodoi{10.1093/mnras/sts345}

\bibitem[{{Montalto} {et~al.}(2020){Montalto}, {Borsato}, {Granata},
  {Lacedelli}, {Malavolta}, {Manthopoulou}, {Nardiello}, {Nascimbeni}, \&
  {Piotto}}]{Montalto2020}
{Montalto}, M., {Borsato}, L., {Granata}, V., {et~al.} 2020, \mnras, 498, 1726,
  \dodoi{10.1093/mnras/staa2438}

\bibitem[{{Nardiello} {et~al.}(2019){Nardiello}, {Borsato}, {Piotto},
  {Colombo}, {Manthopoulou}, {Bedin}, {Granata}, {Lacedelli}, {Libralato},
  {Malavolta}, {Montalto}, \& {Nascimbeni}}]{Nardiello2019}
{Nardiello}, D., {Borsato}, L., {Piotto}, G., {et~al.} 2019, \mnras, 490, 3806,
  \dodoi{10.1093/mnras/stz2878}

\bibitem[{{Oelkers} \& {Stassun}(2018)}]{Oelkers2018}
{Oelkers}, R.~J., \& {Stassun}, K.~G. 2018, \aj, 156, 132,
  \dodoi{10.3847/1538-3881/aad68e}

\bibitem[{Oelkers {et~al.}(2016)Oelkers, Stassun, Pepper, {De Lee}, \&
  Paegert}]{Oelkers2016}
Oelkers, R.~J., Stassun, K.~G., Pepper, J., {De Lee}, N., \& Paegert, M. 2016,
  in 2016 IEEE Int. Conf. Big Data (Big Data) (IEEE), 3204--3213,
  \dodoi{10.1109/BigData.2016.7840976}

\bibitem[{Pedregosa {et~al.}(2011)Pedregosa, Varoquaux, Gramfort, Michel,
  Thirion, Grisel, Blondel, Prettenhofer, Weiss, Dubourg, Vanderplas, Passos,
  Cournapeau, Brucher, Perrot, \& Duchesnay}]{scikit-learn}
Pedregosa, F., Varoquaux, G., Gramfort, A., {et~al.} 2011, Journal of Machine
  Learning Research, 12, 2825

\bibitem[{{Pope} {et~al.}(2019){Pope}, {White}, {Farr}, {Yu}, {Greklek-McKeon},
  {Huber}, {Aerts}, {Aigrain}, {Bedding}, {Boyajian}, {Creevey}, \&
  {Hogg}}]{halopope}
{Pope}, B. J.~S., {White}, T.~R., {Farr}, W.~M., {et~al.} 2019, \apjs, 245, 8,
  \dodoi{10.3847/1538-4365/ab3d29}

\bibitem[{{Price-Whelan} {et~al.}(2018){Price-Whelan}, {Sip{\H{o}}cz},
  {G{\"u}nther}, {Lim}, {Crawford}, {Conseil}, {Shupe}, {Craig}, {Dencheva},
  {Ginsburg}, {VanderPlas}, {Bradley}, {P{\'e}rez-Su{\'a}rez}, {de Val-Borro},
  {Paper Contributors}, {Aldcroft}, {Cruz}, {Robitaille}, {Tollerud},
  {Coordination Committee}, {Ardelean}, {Babej}, {Bach}, {Bachetti}, {Bakanov},
  {Bamford}, {Barentsen}, {Barmby}, {Baumbach}, {Berry}, {Biscani}, {Boquien},
  {Bostroem}, {Bouma}, {Brammer}, {Bray}, {Breytenbach}, {Buddelmeijer},
  {Burke}, {Calderone}, {Cano Rodr{\'\i}guez}, {Cara}, {Cardoso}, {Cheedella},
  {Copin}, {Corrales}, {Crichton}, {D{\textquoteright}Avella}, {Deil},
  {Depagne}, {Dietrich}, {Donath}, {Droettboom}, {Earl}, {Erben}, {Fabbro},
  {Ferreira}, {Finethy}, {Fox}, {Garrison}, {Gibbons}, {Goldstein}, {Gommers},
  {Greco}, {Greenfield}, {Groener}, {Grollier}, {Hagen}, {Hirst}, {Homeier},
  {Horton}, {Hosseinzadeh}, {Hu}, {Hunkeler}, {Ivezi{\'c}}, {Jain}, {Jenness},
  {Kanarek}, {Kendrew}, {Kern}, {Kerzendorf}, {Khvalko}, {King}, {Kirkby},
  {Kulkarni}, {Kumar}, {Lee}, {Lenz}, {Littlefair}, {Ma}, {Macleod},
  {Mastropietro}, {McCully}, {Montagnac}, {Morris}, {Mueller}, {Mumford},
  {Muna}, {Murphy}, {Nelson}, {Nguyen}, {Ninan}, {N{\"o}the}, {Ogaz}, {Oh},
  {Parejko}, {Parley}, {Pascual}, {Patil}, {Patil}, {Plunkett}, {Prochaska},
  {Rastogi}, {Reddy Janga}, {Sabater}, {Sakurikar}, {Seifert}, {Sherbert},
  {Sherwood-Taylor}, {Shih}, {Sick}, {Silbiger}, {Singanamalla}, {Singer},
  {Sladen}, {Sooley}, {Sornarajah}, {Streicher}, {Teuben}, {Thomas},
  {Tremblay}, {Turner}, {Terr{\'o}n}, {van Kerkwijk}, {de la Vega}, {Watkins},
  {Weaver}, {Whitmore}, {Woillez}, {Zabalza}, \& {Contributors}}]{astropy:2018}
{Price-Whelan}, A.~M., {Sip{\H{o}}cz}, B.~M., {G{\"u}nther}, H.~M., {et~al.}
  2018, \aj, 156, 123, \dodoi{10.3847/1538-3881/aabc4f}

\bibitem[{Ricker {et~al.}(2014)Ricker, Winn, Vanderspek, Latham, Bakos, Bean,
  Berta-Thompson, Brown, Buchhave, Butler, Butler, Chaplin, Charbonneau,
  Christensen-Dalsgaard, Clampin, Deming, Doty, {De Lee}, Dressing, Dunham,
  Endl, Fressin, Ge, Henning, Holman, Howard, Ida, Jenkins, Jernigan, Johnson,
  Kaltenegger, Kawai, Kjeldsen, Laughlin, Levine, Lin, Lissauer, MacQueen,
  Marcy, McCullough, Morton, Narita, Paegert, Palle, Pepe, Pepper, Quirrenbach,
  Rinehart, Sasselov, Sato, Seager, Sozzetti, Stassun, Sullivan, Szentgyorgyi,
  Torres, Udry, \& Villasenor}]{Ricker2014}
Ricker, G.~R., Winn, J.~N., Vanderspek, R., {et~al.} 2014, J. Astron. Telesc.
  Instruments, Syst., 1, 014003, \dodoi{10.1117/1.JATIS.1.1.014003}

\bibitem[{Skrutskie {et~al.}(2006)Skrutskie, Cutri, Stiening, Weinberg,
  Schneider, Carpenter, Beichman, Capps, Chester, Elias, Huchra, Liebert,
  Lonsdale, Monet, Price, Seitzer, Jarrett, Kirkpatrick, Gizis, Howard, Evans,
  Fowler, Fullmer, Hurt, Light, Kopan, Marsh, McCallon, Tam, Dyk, \&
  Wheelock}]{2mass}
Skrutskie, M.~F., Cutri, R.~M., Stiening, R., {et~al.} 2006, The Astronomical
  Journal, 131, 1163, \dodoi{10.1086/498708}

\bibitem[{Stassun {et~al.}(2018)Stassun, Oelkers, Pepper, Paegert, Lee, Torres,
  Latham, Charpinet, Dressing, Huber, Kane, L{\'{e}}pine, Mann, Muirhead,
  Rojas-Ayala, Silvotti, Fleming, Levine, \& Plavchan}]{Stassun2018}
Stassun, K.~G., Oelkers, R.~J., Pepper, J., {et~al.} 2018, Astron. J., 156,
  102, \dodoi{10.3847/1538-3881/aad050}

\bibitem[{{Stassun} {et~al.}(2019){Stassun}, {Oelkers}, {Paegert}, {Torres},
  {Pepper}, {De Lee}, {Collins}, {Latham}, {Muirhead}, {Chittidi},
  {Rojas-Ayala}, {Fleming}, {Rose}, {Tenenbaum}, {Ting}, {Kane}, {Barclay},
  {Bean}, {Brassuer}, {Charbonneau}, {Ge}, {Lissauer}, {Mann}, {McLean},
  {Mullally}, {Narita}, {Plavchan}, {Ricker}, {Sasselov}, {Seager}, {Sharma},
  {Shiao}, {Sozzetti}, {Stello}, {Vanderspek}, {Wallace}, \&
  {Winn}}]{Stassun2019}
{Stassun}, K.~G., {Oelkers}, R.~J., {Paegert}, M., {et~al.} 2019, \aj, 158,
  138, \dodoi{10.3847/1538-3881/ab3467}

\bibitem[{{Sullivan} {et~al.}(2015){Sullivan}, {Winn}, {Berta-Thompson},
  {Charbonneau}, {Deming}, {Dressing}, {Latham}, {Levine}, {McCullough},
  {Morton}, {Ricker}, {Vanderspek}, \& {Woods}}]{Sullivan2015}
{Sullivan}, P.~W., {Winn}, J.~N., {Berta-Thompson}, Z.~K., {et~al.} 2015, \apj,
  809, 77, \dodoi{10.1088/0004-637X/809/1/77}

\bibitem[{Twicken {et~al.}(2020)Twicken, Caldwell, Jenkins, Tenenbaum, Smith,
  Wohler, Rose, Ting, Vanderspek, Morgan, Rudat, Fausnaugh, Fleming, \&
  Quintana}]{Twicken2020}
Twicken, J.~D., Caldwell, D.~A., Jenkins, J.~M., {et~al.} 2020, {TESS Science
  Data Products Description Document EXP-TESS-ARC-ICD-0014 Rev F},  NASA Ames
  Research Center.
\newblock
  \url{https://archive.stsci.edu/missions/tess/doc/EXP-TESS-ARC-ICD-TM-0014-Rev-F.pdf}

\bibitem[{van~der Walt {et~al.}(2014)van~der Walt, {S}ch\"onberger,
  {Nunez-Iglesias}, {B}oulogne, {W}arner, {Y}ager, {G}ouillart, {Y}u, \& the
  scikit-image contributors}]{scikit-image}
van~der Walt, S., {S}ch\"onberger, J.~L., {Nunez-Iglesias}, J., {et~al.} 2014,
  PeerJ, 2, e453, \dodoi{10.7717/peerj.453}

\bibitem[{{Vanderburg} \& {Johnson}(2014)}]{Vanderburg2014}
{Vanderburg}, A., \& {Johnson}, J.~A. 2014, \pasp, 126, 948,
  \dodoi{10.1086/678764}

\bibitem[{{Vanderspek} {et~al.}(2018){Vanderspek}, {Doty}, {Fausnaugh},
  {Villasenor}, {Jenkins}, {Berta-Thompson}, {Burke}, \&
  {Ricker}}]{TESSInstrumentHandbook}
{Vanderspek}, R., {Doty}, J.~P., {Fausnaugh}, M., {et~al.} 2018, {TESS
  Instrument Handbook}, Tech. rep., Kavli Institute for Astrophysics and Space
  Science, Massachusetts Institute of Technology.
\newblock
  \url{https://archive.stsci.edu/missions/tess/doc/TESS_Instrument_Handbook_v0.1.pdf}

\bibitem[{{Virtanen} {et~al.}(2020){Virtanen}, {Gommers}, {Oliphant},
  {Haberland}, {Reddy}, {Cournapeau}, {Burovski}, {Peterson}, {Weckesser},
  {Bright}, {van der Walt}, {Brett}, {Wilson}, {Jarrod Millman}, {Mayorov},
  {Nelson}, {Jones}, {Kern}, {Larson}, {Carey}, {Polat}, {Feng}, {Moore}, {Vand
  erPlas}, {Laxalde}, {Perktold}, {Cimrman}, {Henriksen}, {Quintero}, {Harris},
  {Archibald}, {Ribeiro}, {Pedregosa}, {van Mulbregt}, \&
  {Contributors}}]{scipy}
{Virtanen}, P., {Gommers}, R., {Oliphant}, T.~E., {et~al.} 2020, Nature
  Methods, 17, 261, \dodoi{10.1038/s41592-019-0686-2}

\bibitem[{{von Essen} {et~al.}(2020){von Essen}, {Lund}, {Handberg}, {Sosa},
  {Gadeberg}, {Kjeldsen}, {Vanderspek}, {Mortensen}, {Mallonn}, {Mammana},
  {Morgan}, {Villase{\~n}or}, {Fausnaugh}, \& {Ricker}}]{tess_time}
{von Essen}, C., {Lund}, M.~N., {Handberg}, R., {et~al.} 2020, \aj, 160, 34,
  \dodoi{10.3847/1538-3881/ab93dd}

\bibitem[{{White} {et~al.}(2017){White}, {Pope}, {Antoci}, {P{\'a}pics},
  {Aerts}, {Gies}, {Gordon}, {Huber}, {Schaefer}, {Aigrain}, {Albrecht},
  {Barclay}, {Barentsen}, {Beck}, {Bedding}, {Fredslund Andersen}, {Grundahl},
  {Howell}, {Ireland}, {Murphy}, {Nielsen}, {Silva Aguirre}, \&
  {Tuthill}}]{halo}
{White}, T.~R., {Pope}, B.~J.~S., {Antoci}, V., {et~al.} 2017, \mnras, 471,
  2882, \dodoi{10.1093/mnras/stx1050}

\end{thebibliography}
\end{document}